\documentclass[aps,pra,10pt,twocolumn,notitlepage,nofootinbib]{revtex4-1}

\usepackage[utf8]{inputenc}
\usepackage[T1]{fontenc}

\usepackage{times}

\usepackage{verbatim}
\usepackage{mathtools}
\usepackage{xcolor}
\usepackage{amsmath,amsthm,amssymb}
\usepackage{dsfont}
\usepackage[cmintegrals]{newtxmath}
\usepackage{bm}
\usepackage{bbm}
\usepackage{graphicx}

\usepackage[scr=dutchcal]{mathalfa}
\DeclareMathAlphabet{\matheuler}{T1}{cmtt}{m}{n}

\definecolor{myurlcolor}{rgb}{0,0,0.7}
\definecolor{myrefcolor}{rgb}{0.8,0,0}
\usepackage[unicode=true,pdfusetitle, bookmarks=false,bookmarksnumbered=false,
bookmarksopen=false, breaklinks=false,pdfborder={0 0 0},backref=false,
colorlinks=true, linkcolor=myrefcolor,citecolor=myurlcolor,urlcolor=myurlcolor]{hyperref}

\DeclareGraphicsExtensions{.png,.pdf,.eps}
\graphicspath{ {./figs/} }

\usepackage{array}
\newcolumntype{M}[1]{>{\centering\arraybackslash}m{#1}}
\newcolumntype{N}{@{}m{0pt}@{}}

\AtBeginDocument{%
    \newwrite\bibnotes
    \def\bibnotesext{Notes.bib}
    \immediate\openout\bibnotes=\jobname\bibnotesext
    \immediate\write\bibnotes{@CONTROL{REVTEX41Control}}
    \immediate\write\bibnotes{@CONTROL{%
    apsrev41Control,author="08",editor="1",pages="1",title="0",year="1"}}
     \if@filesw
     \immediate\write\@auxout{\string\citation{apsrev41Control}}%
    \fi
}%

\makeatletter

\def\dd{\mathrm d}
\def\ii{\mathrm i}
\def\ee{\mathrm e}
\newcommand{\1}{\openone}
\newcommand{\trm}[1]{\textrm{#1}}

\newcommand{\ket}[1]{\left|#1\right\rangle}
\newcommand{\bra}[1]{\left\langle #1\right|}

\newcommand{\ketbra}[2]{|#1\rangle\langle#2|}
\newcommand{\proj}[1]{\ket{#1}\!\bra{#1}}

\newcommand{\tr}[1]{\mathrm{Tr}\!\left\{#1\right\}}

\newcommand{\vV}[1]{\mathbf{#1}}

\newcommand{\m}[1]{\matheuler{#1}}

\def\dD{\mathcal{D}}
\def\tT{\mathcal{T}}

\def\cMq{\mathcal{M}_\mathrm{Q}}

\def\param{\varphi}
\def\paramV{{\boldsymbol{\param}}}
\def\dparamV{{\boldsymbol{\delta\param}}}

\def\g{\mathscr{g}}
\def\gM{\matheuler{g}}
\def\gMp{\mathfrak{g}}

\def\FM{\mathfrak{F}}
\def\FI{\mathcal{F}}
\newcommand{\QFM}[1]{\FM_{#1}}
\newcommand{\QFIg}[1]{\FI_{#1}}
\def\QFIM{\FM_\textrm{Q}}
\def\QFMR{\FM_\textrm{R}}
\def\QFI{\FI_\textrm{Q}}
\def\QFIR{\FI_\textrm{R}}
\def\WYD{\mathfrak{I}}
\def\KM{\mathfrak{J}}
\def\KMI{\mathcal{J}}

\newcommand{\RR}{\mathbb{R}}

\newcommand{\BE}{\begin{equation}}
\newcommand{\EE}{\end{equation}}
\newcommand{\BEA}{\begin{eqnarray}}
\newcommand{\EEA}{\end{eqnarray}}

\def \cL{\mathcal L}
\def \cH{\mathcal H}

\def \cM{\mathcal M}
\def \cB{\mathcal B}

\newcommand{\eref}[1]{(\ref{#1})}
\newcommand{\eqnref}[1]{Eq.~(\ref{#1})}
\newcommand{\eqnsref}[2]{Eqs.~(\ref{#1}) and (\ref{#2})}
\newcommand{\figref}[1]{Fig.~\ref{#1}}
\newcommand{\tabref}[1]{Table~\ref{#1}}
\newcommand{\secref}[1]{Sec.~\ref{#1}}

\newcommand{\citeref}[1]{Ref.~\cite{#1}}

\makeatother

\begin{document}

\title{Geometric approach to quantum statistical inference}

\author{Marcin Jarzyna}
\email{m.jarzyna@cent.uw.edu.pl}
\affiliation{Centre for Quantum Optical Technologies, Centre of New Technologies, University of Warsaw, Banacha 2c, 02-097 Warsaw, Poland}

\author{Jan Ko\l{}ody\'{n}ski}
\email{j.kolodynski@cent.uw.edu.pl}
\affiliation{Centre for Quantum Optical Technologies, Centre of New Technologies, University of Warsaw, Banacha 2c, 02-097 Warsaw, Poland}

\begin{abstract}
We study quantum statistical inference tasks of hypothesis testing and their canonical variations, in order to review relations between their corresponding figures of merit---measures of statistical distance---and demonstrate the crucial differences which
arise in the quantum regime in contrast to the classical setting. In our analysis, we primarily focus on the geometric approach to data inference problems, within which the aforementioned measures can be neatly interpreted as particular forms of divergences that quantify distances in the space of probability distributions or, when dealing with quantum systems, of density matrices. Moreover, with help of the standard language of Riemannian geometry we identify both the metrics such divergences must induce and the relations such metrics must then naturally inherit. Finally, we discuss exemplary applications of such a geometric approach to problems of quantum parameter estimation, ``speed limits'' and thermodynamics.
\end{abstract}

\maketitle

\section{Introduction}
%
\emph{Quantum statistical inference} is a field lying at the intersection of quantum information theory and statistics~\cite{Helstrom1976,Holevo2001,Brandorff-Nielsen2003,Hayashi2005a}. Its main goal is to provide tools that allow for an efficient assessment of \emph{data} in order to extract desired properties of its underlying probability distribution---which crucially describes outcomes of \emph{measurements} performed on a single or multiple \emph{quantum} systems. As the aim of any experiment is to prove (or disprove) a hypothesized theoretical model, whenever a quantum mechanical description is required, quantum statistical inference methods become essential for making any scientific claims. Overall, the key is always to perform inference tasks most efficiently, in particular, obtaining as much information as possible from a given dataset, e.g., maximizing precision or an appropriate information measure, or minimizing the error or its probability. However, being tailored to problems embedded into the quantum realm, another important role of quantum statistical inference is to quantify and study the advantage of features that only quantum systems and measurements may possess. On one hand, a quantum system may exhibit \emph{entanglement}~\cite{Horodecki2009} in between its constituents, which after being measured can then yield correlations stronger than the classical theory allows for~\cite{Bell1964}. On the other hand, a \emph{collective measurement} performed on multiple system copies can uncover, thanks again to its inbuilt quantum correlations, system properties that are classically inaccessible~\cite{Massar1995,Bennett1999a}. 

In fact, the above aspects of a quantum theory turn out to be vital for the tasks of \emph{quantum parameter estimation} and \emph{quantum hypothesis testing}, respectively. By using entangled particles as probes, the error in estimating a parameter encoded independently onto each of them can decrease even as $\propto1/N$ with the particle number $N$, attaining the so-called Heisenberg limit~\cite{Giovannetti2006, Giovannetti2011, Lee2002}---as opposed to the best classical scaling of $\propto1/\sqrt{N}$ imposed by the central limit theorem~\cite{Pezze2009}. In contrast, in quantum hypothesis testing, e.g.~in discrimination problems, it is the ability to perform collective measurements on many system copies, which allows the probability of incorrect decision to asymptotically vanish with a rate that is unattainable by classical means~\cite{Audenaert2007,Audenaert2008,Nussbaum2009,Ogawa2000}. Such observations have led to booming research in quantum metrology~\cite{Giovannetti2006} and sensing~\cite{Degen2017} at core of which lies the quantum estimation theory, as well as other information theoretical tasks such as private communication over quantum channels~\cite{WIlde2017} or quantum key distribution~\cite{renner_security_2005}, whose security and performance largely rely on proofs involving hypothesis testing problems.   

In this work, the motivation is to study figures of merit that arise in quantum statistical inference tasks and establish important relations between them. In order to achieve this, we approach the inference problems with help of the \emph{information geometry}~\cite{Amari1985,Amari2000}, which allows us to directly generalise the geometric picture of a probability space that in the quantum case naturally inherits the structures of quantum states and measurements. In particular, we use these notions to study the distinguishability problem of quantum states that lie very close to one another, for which we show that figures of merit arising in hypothesis testing naturally reduce to quantities emerging in the complimentary estimation and information theoretical tasks. As a result, we are able to provide alternative proofs of properties for the geometrically motivated quantum measures that are then naturally inherited from the geometric structure, e.g.~convexity, additivity or monotonicity of the quantum Fisher information (QFI) and other quantum metrics. Although we gather and combine results previously established in the literature, we obtain this way also novel relations connecting quantum distance measures, e.g.~a lower bound on the QFI obtained with help of the infinitesimal expansion of the trace distance. Finally, we demonstrate that the geometrically motivated relations can bring insights into problems of \emph{quantum parameter estimation, speed limits} and \emph{thermodynamics}. For instance, we identify in the last case which quantum metric is naturally responsible for the deviation from the exact equality in the familiar Clausius inequality, which governs all thermodynamic processes.
\begin{figure*}[t!]
\centering
\includegraphics[width=0.9\textwidth]{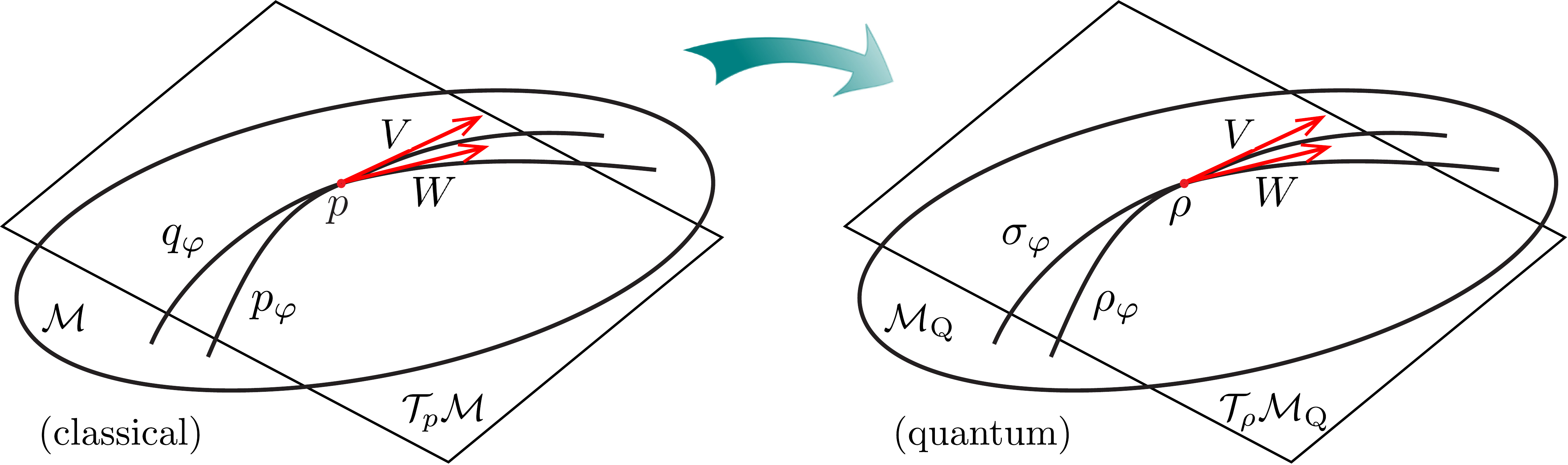}
\caption{{\bf Geometric representation of a \emph{classical} statistical manifold $\cM$ and its \emph{quantum} generalisation $\cMq$}, with the latter containing density matrices rather than probability distributions. The \emph{tangent space} $\tT_{p}\cM$ defined at a given point (probability distribution) $p$ in $\cM$ naturally translates onto 
$\tT_{\rho}\cMq$ defined in a similar manner on $\cMq$ for any point (density matrix) $\rho$. In particular, a tangent space at $p\in\cM$ (or $\rho\in\cMq$) is spanned by all tangent vectors of curves passing through that point, e.g.~vectors $V$ and $W$ tangent to curves $p_{\param}$ and $q_{\param}$ (or $\rho_{\param}$ and $\sigma_{\param}$), respectively, at their crossing-point $p$ (or $\rho$).}
\label{fig:stat_man}
\end{figure*}
%

\section{Classical statistical inference}
\label{sec:cl_stat_inf}
%
We start by introducing the notion of \emph{statistical distances}, their interrelations, as well as conditions one must impose when requiring them to represent valid metrics or divergences in the space of probability distributions. Such a space is geometrically represented  by the \emph{statistical manifold} depicted in \figref{fig:stat_man} that carries natural structures of Riemannian geometry, which, as later discussed, allow to apply tools of \emph{information geometry} in solving statistical inference tasks.

\subsection{Distance measures between probability distributions}
\label{subsec:dist_meas_cl}

\subsubsection{Statistical distances and divergences}
\label{subsec:cl_dist_div}
The purpose of a \emph{statistical distance} is to measure discrepancy between statistical objects, e.g.:~probability distributions, random variables, or samples. Geometrically, any two \emph{probability distributions} (PDs) $p$ and $q$ correspond to two points lying on the (classical) \emph{statistical manifold} $\cM$---the space composed of all legitimate PDs shown schematically in \figref{fig:stat_man}. The distance between them must be generally represented by some non-negative function  $D:\cM\times\cM\to\RR_+$ such that $D\!\left[p,q\right]\ge0$. In this work we would like to consider statistical distances that lead to the intuitive notion of an ``overlap'' between PDs. That is why, we will avoid, for instance, the so-called transport distances%
\footnote{Let us note, however, that these arise naturally in constructions of uncertainty relations for quantum measurements~\cite{Busch2014}, which are beyond the scope of this work.} that instead quantify the ``cost of transforming'' one PD into another~\cite{villani2008optimal}.

A statistical distance $D\!\left[p,q\right]\ge0$ is formally termed a \emph{metric} if:~(\emph{i}) it is `symmetric', $D\!\left[p,q\right]=D\!\left[q,p\right]$;~(\emph{ii}) satisfies the `identity of indiscernibles', $D\!\left[p,q\right]=0\Leftrightarrow p=q$;~and (\emph{iii}) fulfils the `triangle inequality', $D\!\left[p,q\right]+D\!\left[q,r\right]\geq D\!\left[p,r\right]$. However, as we will often use the term ``metric'' referring to its definition in differential geometry, from now on we will call all non-negative functions satisfying conditions (\emph{i}-\emph{iii}) simply as \emph{distances}. On the other hand, as shortly shown in \secref{subsec:Hypothesis-testing}, the above conditions often turn out to be too demanding when seeking statistical distances with a clear operational motivation. That is why, we will call a function $D\!\left[p,q\right]\ge0$ a \emph{divergence} if it is ``almost'' a distance, i.e.~it does not satisfy all of the conditions (\emph{i}-\emph{iii})%
\footnote{Formally, by relaxing condition (\emph{i}), (\emph{ii}) or (\emph{iii}) one arrives at quasi-, pseudo- or semi-metrics, respectively.}. Instead, unless stated otherwise, we will require a divergence to satisfy the `identity of indiscernibles' (\emph{ii}), while not insisting on its `symmetry' (\emph{i}) nor the `triangle inequality' (\emph{iii}).

\subsubsection{$f$-divergences and their monotonicity}
\label{subsec:f-divs}
A particularly useful class of divergences are the so-called $f$\emph{-divergences}~\cite{Ali1966,Csiszar1967}. For any given \emph{convex} function $f:\RR_{+}\to\RR_+$ such that $f(1)=0$, an $f$-divergence is defined as~\cite{Ali1966,Csiszar1967,Liese2006,Sason2016}:
\begin{equation}
D_{f}\!\left[p,q\right]:=\sum_{i}\,p_{i}\,f\!\left(\frac{q_{i}}{p_{i}}\right),
\label{eq:f_div}
\end{equation}
where we have assumed the PDs $p$ and $q$ to be discrete%
\footnote{For simplicity, throughout this work we primarily consider discrete PDs. This is sufficient for our purposes, as in the later sections we deal with quantum states and measurements defined in \emph{finite-dimensional} Hilbert spaces only.}, so that $\sum_{i}p_{i}=1$ and $p_{i}\geq0$ for all the values, $i=0,1,2,\dots$, the corresponding random variable takes (and similarly for $q$). As the form \eref{eq:f_div} does not assure $f$-divergences to be symmetric, we avoid referring to them generally as distances unless all the above conditions (\emph{i}-\emph{iii}) are satisfied.

Importantly, it follows naturally from the convexity of $f$ in \eqnref{eq:f_div} that any $f$-divergence must be \emph{monotonic}~\cite{zakai1975generalization}, i.e.~fulfill the \emph{data-processing inequality} $D_{f}\!\left[p',q'\right]\leq D_{f}\!\left[p,q\right]$ where $p'=\m{S}p$ ($q'=\m{S}q$)%
\footnote{Throughout, we use the same notation for PDs when written in vector form, so that e.g.~$p'=\m{S}p\;\Leftrightarrow\;\forall_i\!:\,p'_i=\sum_j \m{S}_{ij}p_j$.}
is the linearly transformed PD $p$ ($q$)  after action of any \emph{stochastic map} $\m{S}:\cM\to\cM$. Interpreting discrete PDs $p'$ and $p$ as non-negative and normalized vectors, the map $\m{S}$ defines a left-stochastic transition matrix $\m{S}_{ij}$ with non-negative entries ($\forall_{i,j}\!:\,\m{S}_{ij}\ge0$) and columns summing up to $1$ ($\forall_j\!:\,\sum_{i}\m{S}_{ij}=1$)~\cite{gagniuc2017markov}. In fact, one can interpret columns of a transition matrix as vectors of a conditional probability, $\m{S}_{\cdot j}=p(\cdot|j)$, which effectively randomises the input probability vector. Consequently, as under a randomization PDs are expected to approach one another, a valid distance measure on the statistical manifold should be non-increasing under the action of any stochastic map---as assured by the monotonicity of any $f$-divergence \eref{eq:f_div}.

Of course, not every $D:\cM\times\cM\to\RR_+$ that is monotonic under stochastic maps constitutes an $f$-divergence. A counterexample may be easily constructed by composing an $f$-divergence with any non-decreasing function $g:\RR_+\to\RR_+$, such that $g(D_f[\cdot,\cdot])$ is then trivially monotonic but cannot be written as in \eqnref{eq:f_div}. Still, due to the freedom in choosing an arbitrary convex function in \eqnref{eq:f_div}, the class of $f$-divergences turns out to cover a broad spectrum of statistical distances used in data processing and inference. However, as discussed later in \secref{subsec:FI_metric}, surprisingly all $f$-divergences induce (up to a constant factor) via the Chentsov theorem~\cite{Chentsov1978} the same metric on the statistical manifold, despite taking in general different forms depending on $f$ in \eqnref{eq:f_div}.

\subsection{Hypothesis testing tasks}
\label{subsec:Hypothesis-testing}
Let us consider a statistical inference task whose main objective is to correctly identify, based on the outcomes $\vV{x}_n=(x_{1},\dots, x_{n})$ of $n$ independent rounds, which of the two given PDs $p$ and $q$ is the one governing the experiment---constitutes the PD according to which the random variable $X$ describing a single outcome is distributed. The inference procedure can then be fully described by a \emph{decision function} $\dD:X^n\to\{0,1\}$, so that if $\dD(\vV{x}_n)=0$ one concludes $p$ to be the correct PD, while for $\dD(\vV{x}_n)=1$ the PD $q$ is accepted. Unless one of the PDs can be trivially excluded for a given dataset $\vV{x}_n$, one must then minimise the \emph{probability of error} over all possible decision functions in order to find the optimal strategy. 

Such a problem is an instance of a binary \emph{hypothesis testing} (HT) task in which one generally differentiates between the \emph{null} $H_0$ and \emph{alternative} $H_1$ hypotheses of which the first one is given priority---being assumed true prior to collecting any data~\cite{Cover1991}. In particular, for the above problem $H_0$ corresponds simply to the statement:~``$p$ is the true PD'';~while $H_1$ to the opposite:~``$q$ is the true PD''. Within the general approach to HT one defines two types of errors:~\emph{type-I errors} (``false positives'') -- occurring when $H_0$ is rejected based on the data despite actually holding true;~and \emph{type-II errors} (``false negatives'') -- arising when $H_0$ is maintained although the data has actually been generated in accordance with $H_1$. For our problem of distinguishing two PDs $p$ and $q$, the type-I errors are dictated by the conditional probability $P_n(q|p)$ of selecting $q$ as the underlying distribution while the observed sequence $\vV{x}_n$ came actually from $p$. In analogy, the type-II errors occur 
when $p$ is claimed to be behind $\vV{x}_n$ instead of $q$ that is true, i.e.~according to $P_n(p|q)$ which has a similar interpretation to $P_n(q|p)$ with the roles of $p$ and $q$ reversed.

In \emph{symmetric hypothesis testing} (sHT) one treats both error-types at equal grounds, while being interested in minimising the \emph{average error probability}:
\begin{equation}
p_n^\trm{err}:=\pi_p\,P_{n}(q|p)+\pi_q\,P_{n}(p|q),
\label{eq:error_prob_sHT}
\end{equation}
where $\pi_{p(q)}$ are the \emph{a priori} probabilities of $p(q)$ being the true PD. In contrast, for the case of \emph{asymmetric hypothesis testing} (aHT) one is rather interested in minimising the probability of either type-I or type-II error, i.e, either $P_{n}(q|p)$ or $P_{n}(p|q)$ separately, and \emph{not} both of them simultaneously. For this reason, sHT and aHT require slightly different treatment that we briefly review below. Nevertheless, note that since a randomization performed on both $p$ and $q$ can only increase the probability of an error, that is, decrease the distinguishability between the PDs, all information measures emerging in HT should be monotonic under stochastic maps.

\subsubsection{Symmetric Hypothesis Testing}
\label{sec:cl_HT_symm}

\begin{figure}[t!]
\centering
\includegraphics[width=0.8\columnwidth]{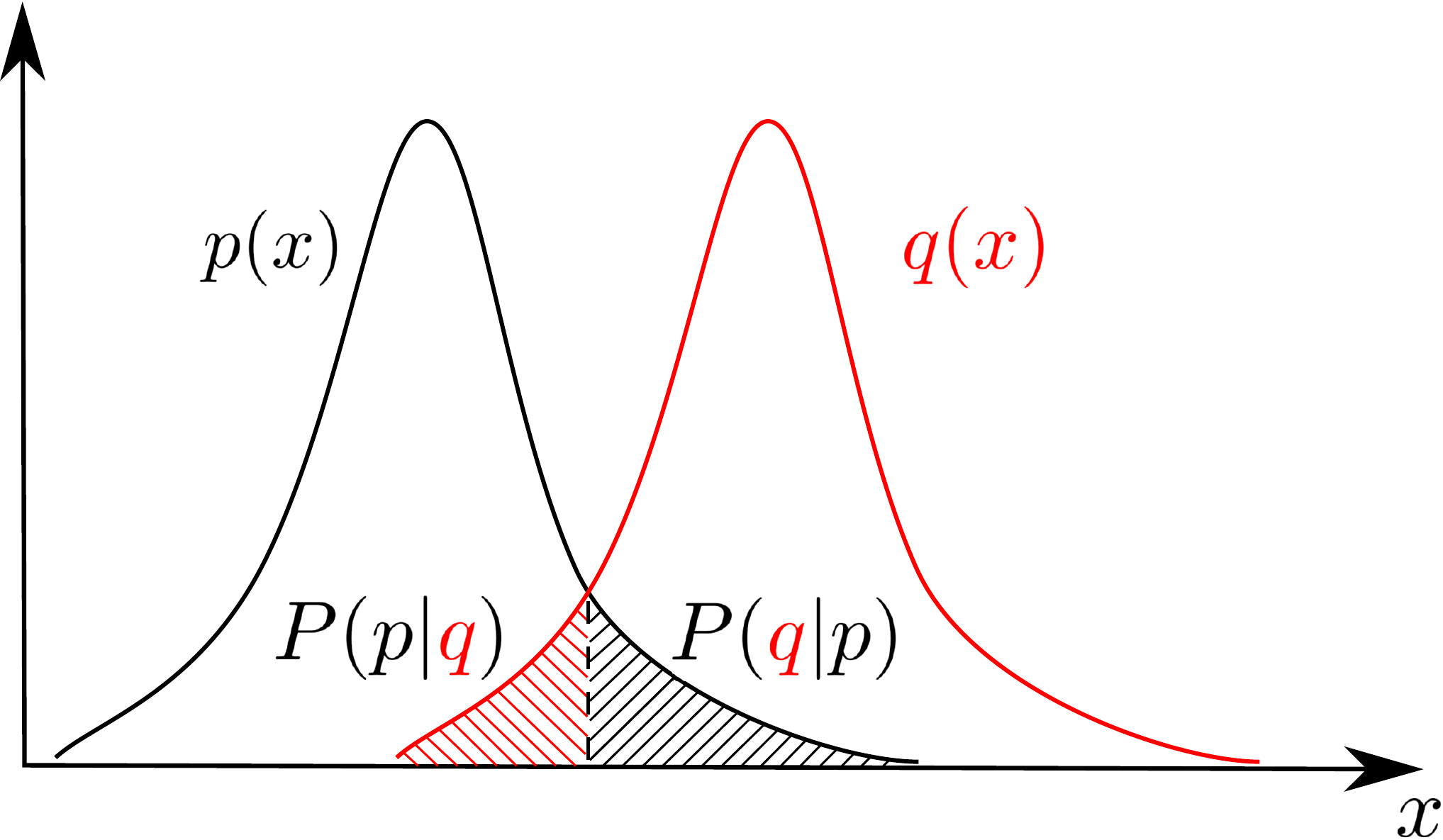}
\caption{\textbf{Symmetric hypothesis testing (sHT) in a single shot}: Distinguishing between continuous probability distributions $p(x)$ and $q(x)$ based on a single ($n=1$) observed measurement outcome $x$. The optimal Neyman-Pearson inference strategy indicates to accept $p(x)$ whenever the measured value $x$ is such that $p(x)\geq q(x)$, and $q(x)$ otherwise. The shaded areas correspond to the type-I error probability (black) and type-II error probability (red). For an unbiased problem $\pi_p=\pi_q=1/2$ the average probability of error, $p_\trm{min}^\trm{err}$ in \eqnref{eq:p_min_err}, corresponds to the total shaded area multiplied by a factor of $1/2$.}
\label{fig:hyp-bin}
\end{figure}

\paragraph{Single-shot scenario ($n=1$)}
Firstly, let us consider the case of distinguishing PDs $p$ and $q$ within the sHT framework based on a single outcome, i.e.~in the \emph{single-shot scenario}. The error probability is then governed by \eqnref{eq:error_prob_sHT} with $n=1$ that must be minimised over all decision functions $\dD:X\to\{0,1\}$ to find the best inference strategy. The optimal $\dD$ corresponds then to the so-called \emph{Neyman-Pearson test}~\cite{Cover1991}, for which the probability of error, $p_{n=1}^\trm{err}$ in \eqnref{eq:error_prob_sHT}, has a natural interpretation depicted in \figref{fig:hyp-bin}. In particular, the Neyman-Pearson strategy instructs one to infer $p(x)$ to be the true distribution if $p(x)\geq q(x)$ for the measured value $x$, and $q(x)$ otherwise. Hence, considering the PDs $p(x)$ and $q(x)$ to be continuous in \figref{fig:hyp-bin} for clarity, the probabilities of type-I and type-II errors correspond then simply to the areas lying under the tails of respective distributions, $P(p|q)=\int_{p(x)\geq q(x)}q(x){\dd x}$ and $P(q|p)=\int_{q(x)\geq p(x)}p(x){\dd x}$, i.e.~the shaded regions marked in \figref{fig:hyp-bin} in red and black, respectively.

On the other hand, substituting the above conditional probabilities into \eqnref{eq:error_prob_sHT} with $n=1$ and using $\pi_p+\pi_q=1$, as well as $\int_{p(x)\ge q(x)}\dots\dd x=1-\int_{q(x)\ge p(x)}\dots\dd x$, one obtains the \emph{minimal average error probability} in the single-shot scenario as
\begin{align}
p_\trm{min}^\trm{err} 
&:=\frac{1}{2}\left(1-\int\left|\pi_p \,p(x)- \pi_q \,q(x)\right|\dd x\right) \nonumber\\
&\equiv
\frac{1}{2}\left(1-\sum_i\left|\pi_p \,p_i- \pi_q \,q_i\right|\right),
\label{eq:err_class}
\end{align}
where we have also written its form for discrete PDs by simply replacing the integral with a sum over a finite number of values. In case the two hypotheses are equally likely and $\pi_p=\pi_q=\frac{1}{2}$, the above expression can be conveniently rewritten as
\begin{equation}
p_\trm{min}^\trm{err}=\frac{1}{2}\left(1-T[p,q]\right)
\label{eq:p_min_err}
\end{equation}
and neatly interpreted as the total shaded area marked in \figref{fig:hyp-bin} divided by two.

In \eqnref{eq:p_min_err} above we have used the common definition of the \emph{Total Variation} (TV) distance: 
\begin{equation}
T[p,q]:=\frac{1}{2}\int\left|p(x)- q(x)\right|\dd x\equiv\frac{1}{2}\sum_{i}\left|p_{i}-q_{i}\right|,
\label{eq:TV}
\end{equation}
applicable to both continuous and discrete PDs. Note that $T[p,q]$ may be termed a distance, as it satisfies all the conditions (\emph{i}-\emph{iii}) stated in \secref{subsec:cl_dist_div}. Nonetheless, it can also be interpreted as an $f$-divergence with a (convex) function $f(t)=\frac{1}{2}|1-t|$ in \eqnref{eq:f_div} with $f(1)=0$. However, as such $f(t)$ is \emph{not} differentiable at $t=1$, the TV distance crucially differs from other commonly used $f$-divergences that are smooth for all $t\in\RR_+$---a point we return to in \secref{subsec:Information-geometry}.

\paragraph{Asymptotic scenario ($n\to\infty$)}
A similar analysis can also be followed for any finite number $n>1$ of independent rounds appearing in \eqnref{eq:error_prob_sHT}, in which case after minimising over all inference strategies one arrives at a direct generalisation of \eqnref{eq:err_class} that in case of discrete PDs reads:
\begin{equation}
p_{n,\trm{min}}^\trm{err}:=
\frac{1}{2}\left(1-\sum_\vV{i}\left|\pi_p \,p_\vV{i}^{(n)}- \pi_q \,q_\vV{i}^{(n)}\right|\right)
\quad\le\quad
\xi[p,q]^n,
\label{eq:chernoff_bound}
\end{equation}
where $p_\vV{i}^{(n)}=\prod_{k=1}^n p_{i_k}$ (and similarly for $q$) denotes now the probability of obtaining a particular sequence $\vV{i}=(i_1,\dots,i_n)$ of outcomes. However, as indicated in \eqnref{eq:chernoff_bound}, the minimal error probability can then be generally upper-limited by the \emph{Chernoff bound}~\cite{Bahadur1960}:
\begin{equation}
\xi[p,q]:=\min_{0\leq\alpha\leq1}\xi_{\alpha}[p,q]
\quad\text{with}\quad
\xi_{\alpha}[p,q]:=\sum_{i}p_{i}^{\alpha}q_{i}^{1-\alpha}.
\label{eq:clCB}
\end{equation}
which is completely independent of the \emph{a priori} distributions $\pi_p$ and $\pi_q$. This is further manifested by $\xi[p,q]=\xi[q,p]$ always being symmetric, in contrast to the \emph{Chernoff coefficients} $\xi_{\alpha}$ defined above (also known as Hellinger integrals \cite{Liese2006}) whose symmetry cannot be guaranteed for all $0\le\alpha\le1$. The only exception is the special case of $\alpha=1/2$ that yields the so-called \emph{Bhattacharyya coefficient} $F[p,q]:=\xi_{1/2}[p,q]=\sum_{i}\sqrt{p_{i}q_{i}}$
being symmetric by its definition.

In general, the Chernoff coefficients $\xi_{\alpha}$ naturally define a family of $f$-divergences called \emph{Hellinger divergences} parametrised similarly by $\alpha$ \cite{jeffreys1946invariant}:
\begin{align}
H_\alpha[p,q]
&:=\frac{1}{1-\alpha}\left(1-\xi_{\alpha}[p,q]\right) =\frac{1}{1-\alpha}\left(1-\sum_{i}p_{i}^{\alpha}q_{i}^{1-\alpha}\right) \nonumber\\
&=\frac{1}{1-\alpha}\sum_{i}p_{i}\left(1-\left(\frac{q_{i}}{p_i}\right)^{1-\alpha}\right),
\label{eq:Hell_div}
\end{align}
which for an extended range of all $0\le\alpha<1$ and $\alpha>1$ consistently constitute examples of \eqnref{eq:f_div} with $f(t)=\frac{1-t^\alpha}{1-\alpha}$ being convex and satisfying $f(1)=0$~\cite{Liese2006,Sason2016}. The special symmetric case of $\alpha=1/2$ leads to the commonly used \emph{(squared) Hellinger distance}%
\footnote{Often the factor of $2$ is omitted, so that the range of squared Hellinger distance is $[0,1]$ rather than $[0,2]$~\cite{Sason2016}.}, $H_{1/2}[p,q]=2(1-F[p,q])$~\cite{Liese2006}, that---as the name suggests---satisfies all (\emph{i}-\emph{iii}) conditions of \secref{subsec:cl_dist_div}.

Although for any small $n\ge1$ the Chernoff bound \eref{eq:chernoff_bound} may not be very tight and, hence, very meaningful, in the \emph{asymptotic scenario} of infinitely many rounds it dictates the ultimate performance in sHT. In particular, as $n\to\infty$ the minimal error probability in \eqnref{eq:chernoff_bound} is guaranteed to follow an exponential decay~\cite{Cover1991}: 
\begin{equation}
p_{n,\trm{min}}^\trm{err}
\quad\underset{n\to\infty}{=}\quad 
\exp\!\left(-n\,C[p,q]+o(n)\right),
\label{eq:as_prop_err_sHT}
\end{equation}
with the exponent being fully determined by the Chernoff bound \eref{eq:clCB}, where
\begin{equation}
C[p,q]:=-\lim_{n\to\infty}\frac{1}{n}\;\ln p_{n,\trm{min}}^\trm{err}=-\ln\xi[p,q]
\label{eq:chernoff_info}
\end{equation}
is often referred to as the \emph{Chernoff information}. Note that the definition \eref{eq:clCB} naturally implies $\xi[p,q]\le\xi_\alpha[p,q]$ for any $0\le\alpha\le1$, so one can always lower-bound the Chernoff information using any Chernoff coefficient $\xi_\alpha$, e.g., by the Bhattacharyya coefficient $F=\xi_{1/2}$ that yields $C[p,q]\ge-\ln F[p,q]$ and is usually the easiest one to compute.

Moreover, for reasons that will soon become clear, it is also convenient to define a family of the so-called \emph{R\'{e}nyi divergences}~\cite{renyi1961measures}:
\begin{align}
D_\alpha[p||q]&:=\frac{1}{\alpha-1}\ln\!\left(1+(\alpha-1)H_{\alpha}[p,q]\right) \label{eq:Renyi_div}\\
&=\frac{1}{\alpha-1}\ln\xi_{\alpha}[p,q]=\frac{1}{\alpha-1}\ln\left(\sum_{i}p_{i}^{\alpha}q_{i}^{1-\alpha}\right),
\nonumber
\end{align}
which, as emphasised by the definition above, are a one-to-one transformation of Hellinger divergences \eref{eq:Hell_div} of the same order $\alpha$~\cite{Sason2016}. Hence, although $D_\alpha[p||q]$ do not generally constitute $f$-divergences \eref{eq:f_div}, they maintain all the properties of $H_\alpha[p,q]$, in particular, condition (\emph{ii}) of \secref{subsec:cl_dist_div} and the monotonicity under stochastic maps (recall the last paragraph of \secref{subsec:f-divs}).

\subsubsection{Asymmetric Hypothesis Testing}
\label{sec:cl_HT_asymm}
In \emph{asymmetric hypothesis testing} (aHT) the aim is to minimise only the probability of one type of errors, either type-I or type-II, while keeping the other fixed. Focussing on, say, type-II errors, this is equivalent to finding optimal inference strategy for which $P_n(p|q)$ in \eqnref{eq:error_prob_sHT} is minimal, while simultaneously assuring that $P_n(q|p)\le\epsilon$ for some $0<\epsilon<1$. Such a problem becomes interesting in the \emph{asymptotic scenario} of infinitely many outcomes, $n\to\infty$, in which similarly to sHT and \eqnref{eq:as_prop_err_sHT} the probability of (one-type) error must decay exponentially as follows~\cite{Cover1991}:
\begin{gather}
P_{n,\trm{min}}(p|q)
\quad\underset{n\to\infty}{=}\quad 
\exp\!\left[-n\,D[p||q]+o(n)\right],
\label{eq:assym_entr}
\end{gather}
where according to the \emph{Stein's lemma} the exponent corresponds to the \emph{relative entropy}, also known as the \emph{Kullback-Leibler divergence}~\cite{Kullback1951}, between PDs $p$ and $q$:
\begin{align}
D[p||q]&\equiv\lim_{\alpha\to1^-}D_\alpha[p||q]\;\equiv\lim_{\alpha\to1^-}H_\alpha[p||q] \nonumber\\
&:=\sum_{i}p_{i}\ln\frac{p_{i}}{q_{i}}=-\sum_{i}p_{i}\ln\frac{q_{i}}{p_{i}},
\label{eq:rel_entr}
\end{align}
which, as emphasised above, can be interpreted as the (one-sided) limiting case of Hellinger \eref{eq:Hell_div} and R\'{e}nyi \eref{eq:Renyi_div} divergences~\cite{Liese2006}, while still constituting an $f$-divergence \eref{eq:f_div} with $f(t)=-\ln t$.

Note that the asymmetry of the relative entropy \eref{eq:rel_entr} is actually a manifestation of the fact that $P_{n,\trm{min}}(q|p)\neq P_{n,\trm{min}}(p|q)$, not even asymptotically. Moreover, if the support of $p$ does \emph{not} lie within the support of $q$, i.e.~$\sup p\nsubseteq\sup q$, and hence $\exists_i\!:\,p_{i}>0\wedge q_i=0$, then $D[p||q]=\infty$. From the operational point of view it must be so, because as soon as the outcome $i$ is observed in the data, which is almost certain as $n\to\infty$, one becomes sure that $p$ is the true PD, so that consistently $\lim_{n\to\infty}P_{n,\trm{min}}(p|q)=0$ in \eqnref{eq:assym_entr} no matter what forms $p$ and $q$ actually take.

\subsection{Information geometry of probability distributions}
\label{subsec:Information-geometry}
In \emph{information geometry}~\cite{Amari1985,Amari2000} the statistical manifold $\cM$ is interpreted as a \emph{Riemannian manifold} that we schematically depict in \figref{fig:stat_man} as a convex oval shape. Hence, not only each point in $\cM$ represents a PD from the probability space, but also $\cM$ carries the structures common to differential geometry such as curvatures, metrics, geodesics or connections. Although a given $\cM$ may generally be non-Euclidean, a coordinate system can always be defined on the manifold via a parametrisation of all PDs:~$p_{\paramV}$ with $\paramV\in\RR^{d}$ and $d=\dim(\cM)$ such that all points in $\cM$ are covered. Then, a one-parameter family $p_{\param}$ of PDs with a smooth $\param:[0,1]\to\cM$ corresponds to a curve in $\cM$, as drawn in \figref{fig:stat_man} together with another family (curve) $q_{\param}$ that coincides with (crosses) $p_{\param}$ at $\param=\param_0$, at which $p_{\param_0}=q_{\param_0}=:p$ is the PD on the intersection.

\subsubsection{Riemannian metrics and geodesics}
\label{subsec:cl_metrics}
A \emph{Riemannian metric} on a manifold $\cM$ is defined as a smooth mapping $\g(p)\!:\tT_{p}\cM\times \tT_{p}\cM\to\RR_+$ between PDs $p\in\cM$ and inner products $\langle\,\cdot,\cdot\,\rangle_{\g(p)}\ge0$ defined for vectors contained in the tangent space $\tT_{p}\cM$ (see \figref{fig:stat_man}) associated with a given point $p$. Moreover, the metric $\g(p)$ at any $p\in\cM$ may be represented as a \emph{matrix} $\gM(p)$ with entries $\gM_{ij}(p):=\langle e_{i},e_{j}\rangle_{\g(p)}$ specified in a fixed orthonormal basis of vectors $\{e_{i}\}_i$ spanning $\tT_{p}\cM$. As a result, the inner product between any two vectors $V,W\in \tT_{p}\cM$ can be conveniently written as $\langle V,W\rangle_{\g(p)}=V^T\gM(p) W=\sum_{ij}\gM(p)_{ij}V_{i}W_{j}$, with the $p$-dependence often dropped. 

Note that a metric is defined to be \emph{additive} on Cartesian products of statistical manifolds for these to describe joint distributions of independent PDs. Consider distinct manifolds $\cM_{1}$ and $\cM_{2}$ with metrics $\g_{1}$ and $\g_{2}$, respectively, and the product manifold $\cM_{12}=\cM_{1}\times\cM_{2}$. Then, given a point $(p_{1},p_{2})\in\cM_{1}\times\cM_{2}$ and any two vectors $V,W\in \tT_{(p_{1},p_{2})}(\cM_{1}\times\cM_{2})\cong\tT_{p_{1}}(\cM_{1})\oplus\tT_{p_{2}}(\cM_{2})$~\cite{isham1999modern}, the metric $\g_{12}$ on $\cM_{12}$ must yield the inner product $\langle V,W\rangle_{\g_{12}{(p_{1},p_{2})}}=\left(\mathbb{P}_{1}V\right)^{T}\gM_{1}(p_1)\left(\mathbb{P}_{1}W\right)+\left(\mathbb{P}_{2}V\right)^{T}\gM_{2}(p_2)\left(\mathbb{P}_{2}W\right)$, where $\mathbb{P}_{1(2)}$ denotes a projection onto $\tT_{p_{1(2)}}\cM_{1(2)}$. In particular, the matrix representation of $\g_{12}$ in any orthonormal basis of the tangent space exhibits a direct-sum structure $\gM_{12}=\gM_{1}\oplus\gM_{2}$, where $\gM_{1(2)}$ is the matrix representation of $\g_{1(2)}$.

Let us emphasise that a metric $\g$ must be independently defined for a given statistical manifold $\cM$. Once specified, however, it defines a special notion of distance between PDs. For an arbitrary curve $\gamma^{(p,q)}\!:[a,b]\ni t\mapsto u\in \cM$ connecting points $p=\gamma^{(p,q)}(a)$ and $q=\gamma^{(p,q)}(b)$ in $\cM$, the tangent vectors along the curve read $\dot\gamma(t):=\left.\frac{\dd \gamma^{(p,q)}}{\dd t}\right|_u=\sum_i\dot{\gamma}_i e_i$ given a basis $\{e_i\}_i$ for each $\tT_u\cM$. As a result, one may define the length of the curve as $|\gamma^{(p,q)}|:=\int_{a}^{b}\sqrt{\sum_{ij}\gM_{ij}\dot{\gamma}_{i}\dot{\gamma}_{j}}\,\dd t$ \cite{Amari2000}, intuitively understood as a sum (i.e.~integral) over infinitesimal segments of length $\sqrt{\langle\dot{\gamma}(t),\dot{\gamma}(t)\rangle}_\g\,\dd t$ along the curve. In particular, the \emph{geodesic distance} between any two points $p,q\in\cM$ is then unambiguously specified by the metric $\g$ as the length of the shortest curve connecting them, $D_{\g}[p,q]:=\min_{\gamma^{(p,q)}}|\gamma^{(p,q)}|$, while the corresponding curve constitutes the \emph{geodesic}. As demanded for any distance, $D_{\g}$ by construction satisfies then all (\emph{i}-\emph{iii}) conditions of \secref{subsec:cl_dist_div}.

\subsubsection{Divergence-induced metrics and their monotonicity}
\label{sec:div_metr_class}
As any metric can be interpreted to define infinitesimal translations along curves on $\cM$, any particular distance or more generally divergence must induce a particular metric on the manifold~\cite{Amari2010}. Consider a divergence $D[p,q]$ that satisfies (\emph{ii}) in \secref{subsec:cl_dist_div} and is \emph{smooth} in its both arguments $p$ and $q$. $D$ defines then an inner product and, hence, a metric $\g_D(p)$ for any two vectors $V,W\in \tT_{p}\cM$ at $p\in\cM$ as follows
\begin{equation}
\g_D(p)[V,W]:=\langle V,W\rangle_{\g_D(p)}=V^T\gM_{D}(p)W,
\label{eq:div_ind_metric0}
\end{equation}
where the matrix representation of $\g_D(p)$ for some orthonormal vector basis $\{e_i\}_i$ in $\tT_p\cM$ reads 
\begin{equation}
\gM_D(p)_{ij}:=\langle e_i,e_j\rangle_{\g_D(p)}=-\left.\frac{\partial^{2}}{\partial t\partial s}D\!\left[p+t\,e_{i},p+s\,e_{j}\right]\right|_{t=s=0} \!\!\!.
\label{eq:div_ind_metric1}
\end{equation}
With help of the definition \eref{eq:div_ind_metric1} we may compactly write the leading-order of the Taylor expansion for any divergence $D$ when perturbing (by $\delta p \in \tT_p\cM$ and $\delta p\to0$) a given PD $p\in\cM$ onto any $p+\delta p\in\cM$ such that $\sup(p+\delta p)\subseteq\sup(p)$  as
\begin{align}
D\!\left[p,p+\delta p\right]
&=\frac{1}{2}\,\g_{D}(p)\!\left[\delta p,\delta p\right]+O(\delta p^3) \nonumber\\
&=\frac{1}{2}\,\delta p^T\gM_{D}(p)\,\delta p+O(\delta p^3), 
\label{eq:div_ind_metric}
\end{align}
where the $\Theta(\delta p)$-term is always absent above, as any divergence by definition fulfils the condition (\emph{ii}) of \secref{subsec:dist_meas_cl} assuring $D\!\left[p,p\right]=0$ to be a global minimum. Then, any divergence satisfies also $D\!\left[p,p+\delta p\right]=D\!\left[p+\delta p,p\right]+O(\delta p^3)$, despite not being necessarily symmetric, and yields the same metric no matter whether it is the first or second argument perturbed in \eqnref{eq:div_ind_metric}. Note that, however, if $\sup(p+\delta p)\nsubseteq\sup(p)$ it follows from \eqnref{eq:div_ind_metric1} that $p$ is perturbed in a pathological direction such that the metric is divergent and cannot be correctly defined.

The definition \eref{eq:div_ind_metric} takes an especially appealing form if we introduce a local coordinate system on $\cM$ that parametrises all $p_{\paramV}\in\cM$ by a smoothly varying $d$-dimensional vector $\paramV=(\param_{1},\dots,\param_{d})\in\RR^{d}$. For any $p_{\paramV}$ we may then consider small deviations $\text{\ensuremath{\dparamV}}$ of $\paramV$ and write~\cite{Amari2000}:
\begin{equation}
D\!\left[p_{\paramV},p_{\paramV+\dparamV}\right]
\approx
\frac{1}{2}\,\dparamV^{T}\,\gMp_{D}(p_{\paramV})\!\left[\nabla p_{\paramV},\nabla p_{\paramV}\right]\dparamV,
\label{eq:div_ind_metric_phi}
\end{equation}
with
\begin{equation}
\gMp_{D}(p_{\paramV})_{ij}\!\left[\nabla p_{\paramV},\nabla p_{\paramV}\right]:=\left.\frac{\partial^{2}}{\partial\theta_{i}\partial\theta_{j}}D\!\left[p_{\paramV},p_{\boldsymbol{\theta}}\right]\right|_{\boldsymbol{\theta}=\paramV}.
\label{eq:metric_coordinates}
\end{equation}
In this context, the metric $\gMp_{D}(p_{\paramV})[\nabla p_{\paramV},\nabla p_{\paramV}]$ forms a matrix in the $\paramV$-coordinate system and can be interpreted as a measure of susceptibility of $p_{\paramV}$ to small changes of the vector parameter $\paramV$ with the resulting variations of $p_{\paramV}$ being measured by the divergence $D$. However, one must bear again in mind that if $\sup(p_{\paramV+\boldsymbol{\delta \param}})\nsubseteq\sup(p_\paramV)$, \eqnref{eq:metric_coordinates} yields $\gMp_{D}(p_{\paramV})[\nabla p_{\paramV},\nabla p_{\paramV}]$ divergent and the metric can no longer be defined.

Crucially, Eqs.~(\ref{eq:div_ind_metric}-\ref{eq:metric_coordinates}) provide a simplified proof of a general statement---any \emph{monotonic divergence} must induce a metric that is also \emph{monotonic} under stochastic maps~\cite{corcuera1998characterization}. Considering any stochastic map $\m{S}$ such that $p'=\m{S}p$ and $\delta p' =\m{S}\delta p$, the monotonicity of $D$ implies $D[p',p'+\delta p']\le D[p,p+\delta p]$, so \eqnref{eq:div_ind_metric} then yields $\gM_D(p)\ge \m{S}^T\gM_D(\m{S}p)\m{S}$ up to $O(\delta p^3)$%
\footnote{Throughout, we use the scalar notation also for matrix inequalities, so that $A\ge B$ \emph{iff} $A-B\ge0$ is a non-negative matrix.}. The same argumentation applies to \eqnref{eq:div_ind_metric_phi}, but as the metric in \eqnref{eq:metric_coordinates} is now defined for both $p=p_{\paramV}$ and $p'=\m{S}p_{\paramV}$ in the $\paramV$-coordinate system, one obtains $\gMp_D(p_{\paramV})[\nabla p_{\paramV},\nabla p_{\paramV}]\ge\gMp_D(\m{S}p_{\paramV})[\nabla (\m{S}p_{\paramV}),\nabla(\m{S}p_{\paramV})]$. For compactness, in what follows we will skip the tangent-vector arguments in \eqnref{eq:metric_coordinates} and refer to the corresponding ($\paramV$-coordinate) metric as just $\gMp_{D}(p_{\paramV})$.

Finally, let us emphasise that, although any divergence $D$ unambiguously defines a metric $\g_{D}$ on the statistical manifold, the converse is not true. In particular, a given metric may be induced by multiple divergences---different $D$ can yield the same \emph{Hessian matrix} in \eqnsref{eq:div_ind_metric1}{eq:metric_coordinates}. The only divergence uniquely defined by a metric $\g$ is the geodesic distance $D_{\g}$ (which, if monotonic, must be defined by a monotonic metric by the argument stated above).

\subsection{Uniqueness of the Fisher metric and its properties}
\label{subsec:FI_metric}
The most spectacular fact in information geometry is the \emph{Chentsov theorem}~\cite{Chentsov1978}. The theorem states that \emph{all} Remiannian metrics defined on a given $\cM$ that are monotonic must always correspond to the Fisher metric up to a multiplicative constant~\cite{Campbell1986extended}. This means that properties of the Fisher metric described below are universal, as they apply to any monotonic metric. 

In particular, the $f$-divergences \eref{eq:f_div} introduced in \secref{subsec:f-divs} are monotonic and so must be all the metrics induced by them. Hence, by the virtue of Chentsov theorem, any metric arising from an $f$-divergence must be proportional to the Fisher metric. To verify this, we explicitly compute the metric $\gMp_D$ in \eqnref{eq:metric_coordinates} for a general $f$-divergence, $D_f$, while assuming $f(t)$ in \eqnref{eq:f_div} to be twice-differentiable at $t=1$. We obtain (see also~\cite{Amari2010})
\begin{equation}
\gMp_{D_f}(p_{\paramV})=\ddot{f}(1)\;\FM(p_{\paramV}),
\label{eq:full_Fisher_metric}
\end{equation}
where
\begin{equation}
\FM(p_{\paramV})_{ij}:=
\sum_{x}\frac{1}{p_{\paramV}(x)}\frac{\partial p_{\paramV}(x)}{\partial\param_{i}}\frac{\partial p_{\paramV}(x)}{\partial\param_{j}},
\label{eq:Fisher_metric}
\end{equation}
is, indeed, the \emph{Fisher metric}~\cite{Campbell1986extended} with the exact form of the $f$-divergence \eref{eq:f_div} affecting only the multiplicative factor $\ddot{f}(1)$ of the overall metric $\gMp_{D_f}$ in \eqnref{eq:full_Fisher_metric}. 

Let us focus on the special case when only a curve in the manifold $\cM$ parametrised by a single $\param\in\mathbb{R}$ is considered, see e.g.~$p_\param$ in \figref{fig:stat_man}. Importantly, the Fisher metric \eref{eq:Fisher_metric} takes then a scalar form of the \emph{Fisher Information} (FI) common to estimation theory~\cite{Lehmann1998}:
\begin{equation}
\FI(p_\param):=\sum_{x}\frac{1}{p_\param(x)}\left(\frac{\partial p_\param(x)}{\partial\param}\right)^{2},
\label{eq:FI}
\end{equation}
which geometrically represents the tangent ``velocity in units of $\param$'' along the curve in $\cM$. As a consequence, all the standard properties associated in the literature with the FI~\cite{Lehmann1998}, can be interpreted as being actually inherited from a monotonic metric, in particular, the \emph{additivity} of FI on independent PDs, $\FI(p_\param q_\param)=\FI(p_\param)+\FI(q_\param)$ for independent $p_\param$ and $q_\param$, and the \emph{monotonicity} of FI under stochastic maps, $\FI(\m{S}p_\param)\le\FI(p_\param)$ for any $\m{S}:\cM\to\cM$ (see \secref{sec:div_metr_class}). 

Furthermore, the monotonicity of the Fisher metric \eref{eq:Fisher_metric} assures its \emph{convexity} when evaluated on mixtures of PD, i.e.~$\FM(\lambda p_\paramV+(1-\lambda)q_\paramV)\leq\lambda\FM(p_\paramV)+(1-\lambda)\FM(q_\paramV)$ for any $0\le\lambda\le1$. In order to prove it, let us consider any two PDs $p_\paramV,q_\paramV\in \cM$ and a third distribution $\tilde{p}_\paramV:=\lambda p_\paramV\oplus(1-\lambda)q_\paramV\in \cB_{2}\times\cM$, where the manifold $\cB_{2}$ consists of binary distributions, so that in case of $\tilde{p}_\paramV$ it contains a flag that with probability $\lambda$ (or $1-\lambda$) indicates the actual PD to be $p_\paramV$ (or $q_\paramV$). As $\cB_{2}$, and hence $\lambda$, is not parametrised by the $\paramV$-coordinates, it is easy to verify with \eqnref{eq:Fisher_metric} that $\FM(\tilde{p}_\paramV)=\lambda\FM(p_\paramV)+(1-\lambda)\FM(q_\paramV)$. Now, consider a stochastic map that corresponds to forgetting the flag outcome, i.e.~$\m{S}:\cB_{2}\times\cM\to\cM$ such that $\m{S}\tilde{p}_\paramV=\lambda p_\paramV+(1-\lambda)q_\paramV$. As the Fisher metric constitutes the (only relevant) monotonic metric, it crucially satisfies $\FM(\m{S}\tilde{p}_\paramV)\le\FM(\tilde{p}_\paramV)$ that is equivalent to $\FM(\lambda p_\paramV+(1-\lambda)q_\paramV)\leq\lambda\FM(p_\paramV)+(1-\lambda)\FM(q_\paramV)$, which proves the convexity, as required. Note that convexity of the FI \eref{eq:FI} trivially follows  by restricting to the single-parameter case---without need to resorting to the exact form of $\FI$ in \eqnref{eq:FI}, as in the original proof \cite{Cohen1968}.

\subsection{Metrics induced by divergences}
\label{subsec:divergence_metrics_induced}
Let us return to the binary HT tasks discussed in \secref{subsec:Hypothesis-testing} and consider the case when two PDs being compared, $p_{\paramV}$ and $p_{\paramV+\dparamV}$, are infinitely close to one another. We can then perform the Taylor expansion in $\dparamV$ of relevant quantities describing the behaviour of the average error probability. Considering the asymptotic scenario of HT, we expand the Chernoff coefficients \eref{eq:clCB}:
\begin{equation}
\xi_{\alpha}\!\left[p_{\paramV},p_{\paramV+\dparamV}\right]\approx1-\frac{1}{2}\alpha(1-\alpha)\dparamV^{T}\FM(p_{\paramV})\dparamV,
\label{eq:Chern_FI}
\end{equation}
that arise in sHT, and similarly the relative entropy (Kullback-Leibler divergence):
\begin{equation}
D\!\left[p_{\paramV}||p_{\paramV+\dparamV}\right]\approx\frac{1}{2}\dparamV^{T}\FM(p_{\paramV})\dparamV,
\label{eq:entr_fi_class}
\end{equation}
which determines in  aHT the asymptotic exponent of the one-type error probability in \eqnref{eq:assym_entr}. In both cases, the Fisher metric, $\FM(p_\paramV)$ defined in \eqnref{eq:Fisher_metric}, can be similarly to \eqnref{eq:div_ind_metric_phi} interpreted as the Hessian matrix (up to a constant factor) for each of the vector-valued Taylor expansions.

Note that the Chernoff coefficients $\xi_{\alpha}$ expanded to the second $\dparamV$-order in \eqnref{eq:Chern_FI} have a simple dependence on the $\alpha$-parameter, so that $\min_\alpha \xi_\alpha$ is always attained for $\alpha=\frac{1}{2}$ as $\dparamV\to0$. As a consequence, when expanding the Chernoff bound \eref{eq:clCB}, which is minimised over $\alpha$, it follows directly from \eqnref{eq:Chern_FI} that the relevant expansion must simply read:
\begin{equation}
\xi\!\left[p_{\paramV},p_{\paramV+\dparamV}\right]\approx1-\frac{1}{8}\dparamV^{T}\FM(p_{\paramV})\dparamV.
\label{eq:chern_clas_FI}
\end{equation}

The above expansions allow us to determine explicitly metrics induced by all the divergences (and distances) discussed in \secref{subsec:Hypothesis-testing}, which we summarise in \tabref{tab:div_class}. As expected, in case of each $f$-divergence the induced metric obeys \eqnref{eq:Fisher_metric}---being proportional to the Fisher metric in accordance with the Chentsov theorem. In case of R\'{e}nyi divergences and the Chernoff information, which do not constitute $f$-divergences, we can easily determine their corresponding metrics thanks to \eqnsref{eq:Chern_FI}{eq:chern_clas_FI}, respectively. The important exception is the TV distance \eref{eq:TV}, which despite being interpretable as an $f$-divergence yields $f(t)=\frac{1}{2}|1-t|$ in \eqnref{eq:f_div} that is not differentiable at $t=1$. Hence, it does \emph{not} induce a Riemannian metric on $\cM$.
\begin{table*}[t]
\centering
\begin{tabular}
{ | M{3.75cm} | M{5cm}| M{1.5cm}| M{1cm}| M{3cm}|N }
\hline 
Classical divergences& $D_f\!\left[p,q\right]$ & $f(t)$ & $\ddot{f}(1)$ & Riemannian metric 
&\\[10pt]
\hline 
\hline 
Hellinger \eref{eq:Hell_div} & $H_{\alpha}[p,q]:=\frac{1}{1-\alpha}(1-\xi_{\alpha}[p,q])$ & $\frac{1-t^\alpha}{1-\alpha}$ & $\alpha$ & $\alpha\,\FM$
&\\[8pt]
\hline 
R\'{e}nyi \eref{eq:Renyi_div} & $D_\alpha[p||q]=\frac{1}{\alpha-1}\ln\xi_{\alpha}[p,q]$ & N/A & N/A & $\alpha\,\FM$
&\\[8pt]
\hline 
Relative entropy \eref{eq:rel_entr} & $D[p||q]=\sum_{i}p_{i}\ln\frac{p_{i}}{q_{i}}$ & $-\ln t$ & 1 & $\FM$
&\\[8pt]
\hline 
Chernoff information \eref{eq:chernoff_info} & $C[p,q]=-\ln\!\left(\min_{0\leq\alpha\leq1}\xi_{\alpha}[p,q]\right)$ & N/A & N/A & $\frac{1}{4}\,\FM$
&\\[8pt]
\hline 
Hellinger distance & $H_{1/2}[p,q]=2(1-F[p,q])$ & $2(1-\sqrt{t})$ & $\frac{1}{2}$ & $\frac{1}{2}\,\FM$
&\\[8pt]
\hline
Total variation distance \eref{eq:TV} & $T[p,q]=\frac{1}{2}\sum_{i}\left|p_{i}-q_{i}\right|$ & $\frac{|1-t|}{2}$ & N/A & N/A
&\\[8pt]
\hline
\end{tabular}
~\\~
\caption{Divergences defined by means of Chernoff (and Bhattacharyya) coefficients $\xi_{\alpha}[p,q]=\sum_{i}p_{i}^{\alpha}q_{i}^{1-\alpha}$ (and~$F[p,q]=\xi_{1/2}[p,q]=\sum_{i}\sqrt{p_{i}q_{i}}$) interpreted as $f$-divergences whenever possible, and their associated Riemannian metrics in relation to the Fisher metric $\FM$.}
\label{tab:div_class}
\end{table*}
%

\subsection{Relations between divergences and metrics}
\label{sec:class_relations}
Although in \secref{subsec:Hypothesis-testing} we have introduced different divergences in distinct contexts of HT, majority of them are, in fact, interrelated~\cite{Sason2016}. In what follows, we present only the most common relations, but importantly compute their ``infinitesimal'' versions---what allows us to relate also the metrics induced by each divergence. We on purpose omit below the PD-arguments of the discussed divergences whenever possible, in order to emphasise that the following inequalities constitute functional relations that do not depend on the actual PDs being considered.

\subsubsection{Relating divergences}
Considering first the case of sHT, we relate the TV distance \eref{eq:TV} -- applicable in the single shot scenario;~with Chernoff coefficients \eref{eq:clCB} -- applicable in the scenario of multiple (incl.~infinite) rounds; by the following ``sandwich'' inequality~\cite{Sason2016,Audenaert2014}: 
\begin{equation}
1-\xi_{\alpha}\le T\le\sqrt{1-F^{2}}.
\label{eq:Clas_Fuchs}
\end{equation}
Inspecting definitions of $T$, $\xi_\alpha$ and $F=\xi_{1/2}$ in \eqnsref{eq:TV}{eq:clCB}, note that the left inequality above directly follows from $\sum_{i}\left(1-\left(\frac{p_{i}}{q_{i}}\right)^{\alpha}\right)q_{i}\le\frac{1}{2}\sum_{i}\left|1-\frac{p_{i}}{q_{i}}\right|q_{i}$, as $1-x^{\alpha}\leq\frac{1}{2}|1-x|$ for any $0\leq\alpha\leq1$. The right inequality, on the other hand, between the TV distance and the Bhattacharyya coefficient can be proven as follows:
\begin{align}
T[p,q]&=\frac{1}{2}\sum_{i}\left|p_{i}-q_{i}\right|
\leq
\frac{1}{2}\sum_{i}\left|\sqrt{p_{i}}-\sqrt{q_{i}}\right|\left|\sqrt{p_{i}}+\sqrt{q_{i}}\right|\nonumber\\
&\leq\frac{1}{2}\sqrt{\sum_{i}\left(\sqrt{p_{i}}-\sqrt{q_{i}}\right)^{2}\sum_{j}\left(\sqrt{p_{j}}+\sqrt{q_{j}}\right)^2}\nonumber\\
&=\sqrt{1-F[p,q]^2},
\end{align}
where in the second inequality is of a Cauchy-Schwarz type. Reversing the inequalities in Eq.~(\ref{eq:Clas_Fuchs}), we similarly obtain a "sandwich" inequality for the Chernoff bound \eref{eq:clCB}:
\begin{equation}
\left.\begin{array}{c}
1-T\leq\xi_{\alpha}\\
F=\xi_{1/2}\leq\sqrt{1-T^{2}}
\end{array}\right\} \Longrightarrow1-T\leq\xi\leq\sqrt{1-T^{2}},
\label{eq:class_chern_var}
\end{equation}
which we further minimised over the $\alpha$-parameter, i.e.~$\xi:=\min_\alpha\xi_\alpha$.

In case of aHT we refer to the celebrated \emph{Pinsker inequality}~\cite{Sason2016}:
\begin{equation}
T[p,q]\le\sqrt{\frac{D[p||q]}{2}},
\label{eq:clas_pinsker}
\end{equation}
which relates the relative entropy \eref{eq:rel_entr} -- determining the asymptotic exponent of the asymmetric error probability \eref{eq:assym_entr};~to the TV distance \eref{eq:TV}. Note that \eqnref{eq:clas_pinsker} together with \eqnref{eq:Clas_Fuchs} imply:
\begin{equation}
1-\sqrt{\frac{D[p||q]}{2}}\le\xi[p,q]\leq\xi_{\alpha}[p,q],
\label{eq:Pinsk_chern_clas}
\end{equation}
which connects the behaviours of asymptotic exponents for sHT \eref{eq:as_prop_err_sHT} and aHT \eref{eq:assym_entr}, in particular, the Chernoff bound \eref{eq:clCB} to the relative entropy \eref{eq:rel_entr}. However, acknowledging that the minimal error probability in the asymptotic regime for the symmetric case in \eqnref{eq:as_prop_err_sHT} cannot be smaller than its equivalent in \eqnref{eq:assym_entr} for the asymmetric scenario~\cite{Cover1991}, we may further tighten the above bound based on the Pinsker inequality \eref{eq:Pinsk_chern_clas} as follows:
\begin{align}
&D[p||q]\geq-\ln\xi[p,q]\nonumber\\
&\implies\qquad
\exp(-D[p||q])\le\xi[p,q]\leq\xi_{\alpha}[p,q].
\label{eq:relative_chernoff_class}
\end{align}

\subsubsection{Metric-induced bounds on the TV distance}
\label{subsec:Induced-relations-on}
Although the TV distance \eref{eq:TV} cannot be used to define a Riemannian metric on $\cM$ -- while not constituting a smooth $f$-divergence, see \tabref{tab:div_class};~in order to determine its behavior for PDs infinitely close to one another, i.e.~$p_\paramV$ and $p_{\paramV+\dparamV}$ introduced in \secref{subsec:divergence_metrics_induced}, we can crucially resort to the ``sandwich'' inequality \eref{eq:Clas_Fuchs} involving Chernoff coefficients \eref{eq:clCB}. However, in what follows we restrict for simplicity to the single-parameter case of a curve $p_\param$ parametrised by $\param\in\mathbb{R}$ (see \figref{fig:stat_man}), but all the relations derived below can be straightforwardly generalised to the multiparameter case---for which the scalar inequalities involving the FI \eref{eq:FI}, $\FI$, generalise to matrix inequalities involving the Fisher metric \eref{eq:Fisher_metric}, $\FM$, while the Taylor expansion of the TV distance must be performed in the vector form involving relevant Hessian matrices.

Substituting the (scalar with $\dparamV\equiv\delta\param$) Taylor expansion \eref{eq:Chern_FI} of the Chernoff coefficient, $\xi_\alpha$, into the ``sandwich'' inequality \eref{eq:Clas_Fuchs} and maximizing the lower bound over $0\le\alpha\le1$, we obtain
\begin{equation}
\FI(p_\param)\,\frac{\delta\param^{2}}{8}
\;\le\; 
T[p_\param,p_{\param+\delta\param}]
\;\leq\;
\sqrt{\FI(p_\param)}\,\frac{\delta\param}{2}.
\label{eq:Variation_bound}
\end{equation}
The left and right inequalities above consistently become tight as $\delta\param\to0$, i.e.~as the TV distance converges to zero for identical PDs. However, it is only the upper bound in \eqnref{eq:Variation_bound} that correctly predicts the behavior of $T$ at the $\Theta(\delta\param)$-order (the first derivative). As the Taylor expansion of the TV distance in $\delta\param$ reads $T[p_\param,p_{\param+\delta\param}]=\frac{1}{2}\sum_{x}\left|\frac{\partial p_\param(x)}{\partial\param}\right|\delta\param+O(\delta\param^2)$, it becomes clear that in the $\delta\param\to0$ limit only the right inequality in \eqnref{eq:Variation_bound} yields a non-trivial bound:
\begin{equation}
\left\Vert \frac{\partial p_\param}{\partial\param}\right\Vert_1 ^{2}
\;\leq\;
\FI(p_\param),
\label{eq:clas_derivative}
\end{equation}
where by $||v||_1:=\sum_i|v_i|$ we denote the $\ell_1$ (Manhattan) norm of a vector $v$. Intuitively, \eqnref{eq:clas_derivative} means that the speed of $p_\param$ along the curve in \figref{fig:stat_man}, when evaluated as the squared sum of all the absolute changes in outcome probabilities for a given $\param$, cannot be larger than the corresponding metric along the curve at $p_\param$, i.e.~larger than the FI \eref{eq:FI} introduced in \secref{subsec:FI_metric}.

\section{Quantum statistical inference}
\label{sec:Quantum-statistical-inference}
The laws of quantum mechanics may be viewed, up to some degree, as a generalization of the
classical statistical theory in which instead of PDs one deals with \emph{density matrices}~\cite{Nielsen2000}. Each of the latter kind represents the knowledge available about a given quantum system, i.e.~its \emph{quantum state}, and formally corresponds to a non-negative linear Hermitian operator $\rho\in\cL(\cH)$ with unit trace $\tr{\rho}=1$ acting on some Hilbert space $\cH$. The \emph{pure states} constitute then the extreme points of the density-matrix set corresponding to rank-one matrices $\rho=|\psi\rangle\langle\psi|$, and adequately describe the system if its state is perfectly known (and hence deterministic).

In case the system is composite or, in other words, multipartite, its density matrix describes the overall state of two (or more) subsystems $A$ and $B$, while accounting for the correlations being shared between the two parties. The overall Hilbert space may then be generally decomposed into a tensor product of subsystem Hilbert spaces $\cH_{AB}=\cH_{A}\otimes\cH_{B}$, while the shared state $\rho_{AB}\in\cL(\cH_{A}\otimes\cH_{B})$ does not necessarily can be written in a product form. In particular, the correlations it yields can only be described classically when the state is \emph{separable}, i.e.~decomposable into a convex sum of pure states $\rho_{AB}=\sum_{\ell}p_{\ell}\ket{\psi_\ell}_A\bra{\psi_\ell}\otimes\ket{\phi_\ell}_B\bra{\phi_\ell}$. Otherwise, the state is said to exhibit \emph{entanglement}~\cite{Horodecki2009}---a fundamental feature of quantum mechanics that distinguishes it drastically from the classical realm.

Moreover, unlike classical physics, the correct description of the measurement is as important in the quantum picture. It is so, as, on one hand, the measuring process actively changes the state of a quantum system and, on the other, multipartite (collective) measurements may extract correlations beyond reach of measurements performed separately on each subsystem. On mathematical grounds, the measurement is described with help of a \emph{positive operator value measure} (POVM) $\{\Pi_{i}\}_i$, composed of operators in $\cL(\cH)$ indexed by each outcome $i$ that satisfy $\Pi_{i}\geq0$ and $\sum_{i}\Pi_{i}=\1$. The probability of obtaining an outcome $i$, given the system is in the quantum state $\rho$, is then given by the \emph{Born rule}:~$p_i=\tr{\rho\Pi_{i}}$~\cite{Nielsen2000}.

From the perspective of statistical inference tasks, the quantum setting may still be viewed at the level of PDs just as instances of a classical inference problem, when one possesses a ``deeper'' information about the origin of each PD. This, in particular, should not affect any of the PD-based inference methods and geometric structures discussed in the previous sections. However, in the optimisation of each task and its geometric description one has then the freedom to vary both the quantum state of a system, as well as the measurements being performed---both of which carry now their individual (also geometric) structures unique to quantum mechanics. 

From a different perspective, however, the classical statistical theory may be regarded as a special case of the quantum one in which the density matrices are forced to be diagonal in the basis defined by some projective measurement that is the only one available. As any state $\rho$ is then constrained to yield the same eigenvectors in $\rho=\sum_{n}p_n|n\rangle\langle n|$ for the measurement $\{\Pi_{n}=|n\rangle\langle n|\}_n$ being considered, every possible PD of outcomes $n$ can be associated with a particular vector of state eigenvalues, $p\in\cM$ for $\rho$ above, belonging to just a (classical) statistical manifold---with the quantum structure of states and measurements becoming completely irrelevant.

Still, in the general case, the full information about a quantum state is contained in its density matrix rather than the outcome PD that applies only after a given measurement has been specified. Hence, as the quantum framework must allow to describe the evolution of information under certain processes---similarly as the transformations of PDs are described by stochastic maps $\m{S}:\cM\to\cM$ in \secref{subsec:f-divs}---it must allow to correctly describe generalisations of stochastic maps that transform density matrices. Indeed, a general mapping of a quantum system is formally described by a \emph{quantum map (channel)} $\Lambda: \cL(\cH)\to\cL(\cH)$ that constitutes a completely positive, linear and trace-preserving transformation on $\cL(\cH)$~\cite{Nielsen2000}. For instance, a special and idealistic case is the unitary map $\Lambda[\rho]= U\rho U^{\dagger}$, which describes a noiseless evolution of an isolated quantum system with the unitary operator $U:\cH\to\cH$ constituting, e.g.~the solution to the celebrated Schr\"{o}dinger equation $U=\exp[-\ii \hat{H} t]$ for some physical Hamiltonian $\hat{H}$~\cite{Nielsen2000}. 

However, as any real quantum system will inevitably interact with its environment and/or be destructively affected while being measured, the theory of quantum maps in its full scope constitutes the ``tour de force'' also in quantum statistical inference. Nonetheless, as quantum maps form generalisations of (classical) stochastic maps, they should be regarded as ``quantum randomisations'' applied now on quantum states rather than PDs. Therefore, as one requires all the statistical distance measures---in particular, divergences, see \secref{subsec:f-divs}---to be \emph{monotonic} under action of stochastic maps $\m{S}$, one must similarly ensure their quantum generalisations to decrease under the action of quantum maps $\Lambda$. This important requirement will have a crucial impact on both the HT tasks and the information geometry in the quantum setting, which we discuss in the following sections.

\subsection{Distance measures between quantum states}
\label{subsec:quant_dist_meas}

\subsubsection{Quantum statistical distances and divergences}
Similarly to the classical case of PDs, one may define distance measures on the space of quantum states, i.e.~on the set of positive semidefinite, unit-trace operators acting on some Hilbert space $\cH$. The requirements for a function $D:\cL(\cH)\times\cL(\cH)\to\mathbb{R}_+$ acting now on pairs of quantum states to constitute a valid \emph{quantum distance} are exactly the same as the conditions (\emph{i-iii}) specified in \secref{subsec:cl_dist_div}. Moreover, in the same way that divergences are required to fulfil only the `identity of indiscernibles' (\emph{ii}:~$D[\rho,\sigma]=0\Leftrightarrow\rho=\sigma$), it is not demanded for a \emph{quantum divergence} to neither be `symmetric' (\emph{i}:~$D[\rho,\sigma]=D[\sigma,\rho]$), nor respect the `triangle inequality' (\emph{iii}:~$D[\rho,\sigma]+D[\sigma,\varrho]\ge D[\sigma,\varrho]$), when considering, this time, any quantum states $\rho,\sigma,\varrho\in\cL(\cH)$. 

However, aiming to generalise any (classical) definition of a statistical distance such that it can be unambiguously defined on the space of quantum states, one must ensure its quantum definition to be independent of the measurement choice. This, in particular, can be achieved by performing maximisation over all the POVMs available, i.e.:
\begin{align}   
D[\rho,\sigma]:=&\max_{\{\Pi_i\}_i} D[p,q] \label{eq:cl_dist_to_q_dist}\\
&\;\text{s.t.}\quad
p_i\!=\!\tr{\rho \Pi_i}
\;\&\;
q_i\!=\!\tr{\sigma \Pi_i}\!, \nonumber
\end{align}
whose elements must be non-negative $\Pi_i\ge0$ and satisfy $\sum_i\Pi_i=\1$. In this way, we obtain a natural recipe on how to redefine in the quantum setting a given distance $D[p,q]$ measuring separation between any two PDs $p,q\in\cM$, so that $D[\rho,\sigma]$ in \eqnref{eq:cl_dist_to_q_dist} is now unambiguously specified for a pair of any quantum states $\rho,\sigma\in\cL(\cH)$ and automatically inherits all the conditions (\emph{i-iii}) of \secref{subsec:cl_dist_div} that are originally satisfied by $D[p,q]$~\cite{Nielsen2000}. Furthermore, as motivated in the previous section, one should expect any quantum distance to reproduce its classical equivalent in case the compared quantum states are diagonal in the same basis, i.e.~$\rho=\sum_i p_i \ketbra{i}{i}$ and $\sigma=\sum_i q_i \ketbra{i}{i}$ for which \eqnref{eq:cl_dist_to_q_dist} consistently reads $D[\rho,\sigma]=D[p,q]$ with PDs $p$ and $q$ corresponding to the eigenvalues of $\rho$ and $\sigma$, respectively.

\subsubsection{Quantum f-divergences and their monotonicity under quantum maps}
\label{subsec:quant_f-div}
The classical notion of an $f$-divergence \eref{eq:f_div} introduced in \secref{subsec:f-divs} has been generalised to the quantum realm in the seminal works of Petz~\cite{petz1985quasi,Petz1986}. Considering as in \eqnref{eq:f_div} a \emph{convex} function $f:\RR_+\to\RR_+$ such that $f(1)=0$, its action on for any positive semidefinite operator $\varrho\ge0$  naturally generalises to $f(\varrho):=\sum_i f(\lambda_i)\proj{\chi_i}$ being now specified in the eigenbasis $\varrho=\sum_i \lambda_i\proj{\chi_i}$. Consequently, a \emph{quantum $f$-divergence} between states $\rho,\sigma\in\cL(\cH)$ may be defined for a given convex $f$ as~\cite{Hiai2011}:
\begin{equation}
D_{f}\!\left[\rho,\sigma\right]=\tr{\rho^{1/2}f\!\left(L_{\sigma}R_{\rho^{-1}}\right)\rho^{1/2}},
\label{eq:q_f_div}
\end{equation}
where $L_{\varrho}$ ($R_{\varrho}$) denotes a left- (right-)multiplication superoperator by a state $\varrho\in\cL(\cH)$ such that $L_{\varrho}A:=\varrho A$ ($R_{\varrho}A:=A\varrho$) for any $A\in\cL(\cH)$, while $f\!\left(L_{\sigma}R_{\rho^{-1}}\right):=\sum_i f\!\left(\frac{q_i}{p_i}\right)L_{\psi_i}R_{\phi_i}$ can be evaluated by resorting again to eigendecompositions of $\rho=\sum_ip_i\proj{\phi_i}$ and $\sigma=\sum_iq_i\proj{\psi_i}$ and restricting to the positive subspace of $\rho$ ($p_i>0$) in \eqnref{eq:q_f_div}%
\footnote{By continuity, in case the states $\rho,\sigma$ in \eqnref{eq:q_f_div} are not of full rank, one can equivalently evaluate their quantum $f$-divergence by considering the limit $D_{f}\!\left[\rho,\sigma\right]=\lim_{\epsilon\to0_+}D[\rho+\epsilon\1,\sigma+\epsilon\1]$~\cite{Hiai2017}.}. Note that when $\rho$ and $\sigma$ commute, i.e.~$\forall_i:\ket{\psi_i}=\ket{\phi_i}$, \eqnref{eq:q_f_div} consistently simplifies to the classical definition of an $f$-divergence with $p$ and $q$ in \eqnref{eq:f_div} being specified then by the eigenvalues of $\rho$ and $\sigma$, respectively. 

For the purpose of this paper, let us just state that any quantum $f$-divergence \eref{eq:q_f_div} is \emph{monotonic under the action of general quantum maps}~\cite{Petz1986}. In particular, for all $\rho,\sigma\in\cL(\cH)$ and any quantum (completely positive, trace-preserving) map $\Lambda:\cL(\cH)\to\cL(\cH)$ a general $D_{f}\!\left[\rho,\sigma\right]$ defined according to \eqnref{eq:q_f_div} must always satisfy $D_{f}\!\left[\Lambda[\rho],\Lambda[\sigma]\right]\le D_{f}\!\left[\rho,\sigma\right]$. For an explicit proof of the monotonicity property and its subtle variations beyond the scope of this work, we refer the reader to Refs.~\cite{Hiai2011,Hiai2017,Lesniewski1999,tomamichel2009fully}.

\subsection{Quantum hypothesis testing tasks}
\label{subsec:q_ht}
All the tasks of binary HT introduced in \secref{subsec:Hypothesis-testing} have a natural generalisation in the quantum setting. The crucial difference, however, is that while in classical HT one must decide between two PDs $p,q\in\cM$, in the quantum scenario the aim is to distinguish which of the two quantum states $\rho,\sigma\in\cL(\cH)$ describes a physical system being measured---with the null $H_0$ or alternative $H_1$ hypotheses indicating $\rho$ or $\sigma$ to be the true state, respectively. Still, such a quantum HT task reduces to the classical problem when the measurement (characterised by a POVM $\{\Pi_i\}_i$) is fixed, so that all discussions of \secref{subsec:Hypothesis-testing} then apply with $p_i=\tr{\rho\Pi_{i}} $ and $q_i=\tr{\sigma\Pi_{i}}$.

\begin{figure}[t!]
\centering
\includegraphics[width=0.95\columnwidth]{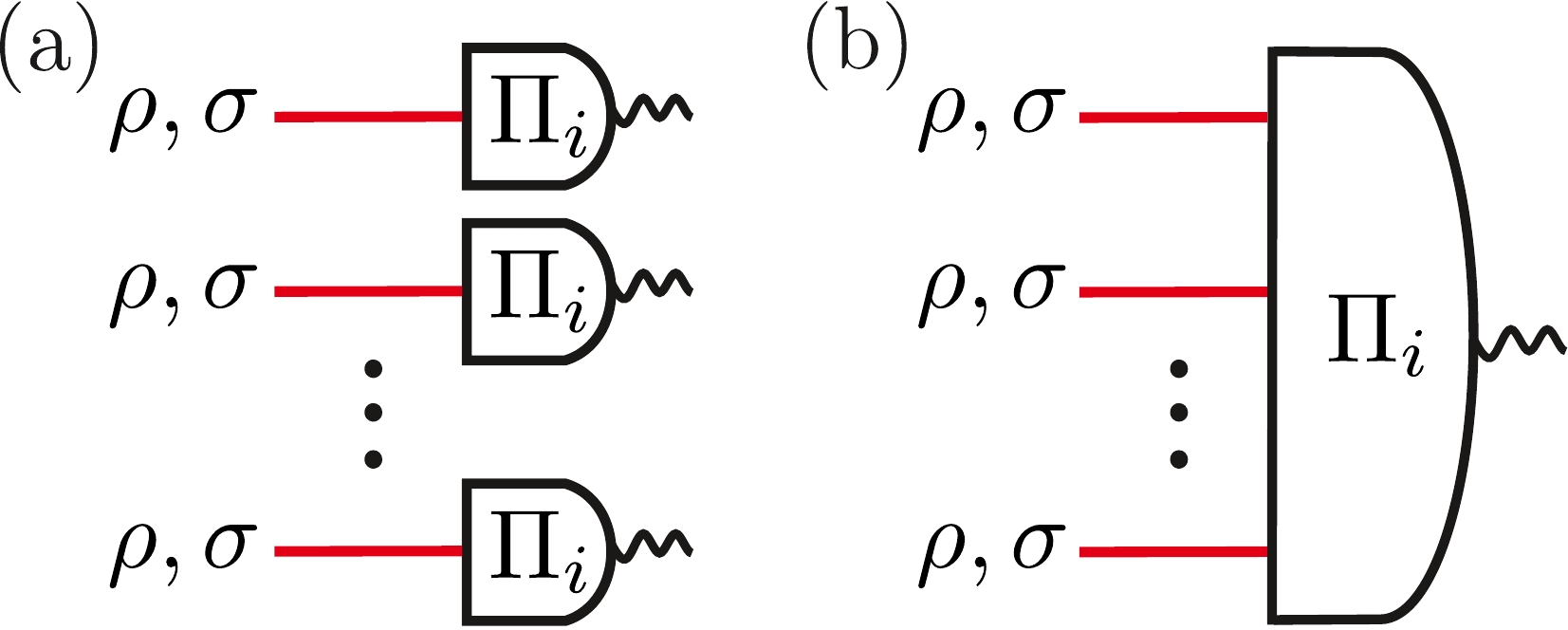}
\caption{\textbf{Quantum hypothesis testing (HT)} of states $\rho$ vs $\sigma$ under \emph{local} (a) and \emph{collective} (b) measurement strategies. In the local case identical POVMs $\{\Pi_i\}_i$ are performed on each system copy, while a collective measurement is represented by a single POVM $\{\Pi_i\}_i$ acting on all the system copies.}
\label{fig:Collective_hypot}
\end{figure}

Similarly to quantum generalisations of statistical distance measures in \eqnref{eq:cl_dist_to_q_dist}, in order to find the optimal strategy in quantum HT one must minimise the average error probability \eref{eq:error_prob_sHT}---either $p_n^\trm{err}$ in sHT or $P_n(p|q)$ in aHT---over all potential measurements (POVMs) that importantly can now be performed collectively on all the $n$ copies of the system, as schematically depicted in \figref{fig:Collective_hypot}(b), so that $p_i=\tr{\rho^{\otimes n}\Pi_{i}}$ and $q_i=\tr{\sigma^{\otimes n}\Pi_{i}}$. The decision strategy, on the other hand, remains captured by the classical HT theory and, hence, its optimisation may be performed following the same procedure discussed in \secref{subsec:Hypothesis-testing}. 

Nonetheless, any quantity dictating the behaviour of the minimal discrimination error in the quantum setting, which we review for sHT followed by aHT below, one should expect to be \emph{monotonic} under the action of quantum maps. This is because a quantum map, which is generally introduced to account for e.g.~disturbance, dissipation, noise or lack of control, when acting on a pair of states can only make them less distinguishable.

%

\subsubsection{Symmetric quantum Hypothesis Testing}
\label{sec:sHT_quantum}

\paragraph{Single-shot scenario ($n=1$)}
Within the framework of sHT and only a single copy of the quantum system available, we may just apply the classical expression \eref{eq:err_class} for the minimal average error probability, $p_\trm{min}^\trm{err}$, given that the measurement $\{\Pi_i\}_i$ is fixed, so that $p_i=\tr{\rho\Pi_{i}} $ and $q_i=\tr{\sigma\Pi_{i}}$ in \eqnref{eq:err_class}. The optimisation over all POVMs ($\forall_i\!:\,\Pi_i\ge0$ and $\sum_i\Pi_i=\1$), however, is then straightforward and yields~\cite{Helstrom1976,Audenaert2007}:
\begin{equation}
p_\trm{min}^\trm{err}:=\frac{1}{2}\left(1-\left\Vert\pi_\rho \,\rho- \pi_\sigma \,\sigma\right\Vert_{1}\right)
\label{eq:err_quant}
\end{equation}
where $\pi_\rho,\pi_\sigma$ are again the \emph{a priori} probabilities of each hypothesis and $||A||_{1}=\tr{\sqrt{A^{\dagger}A}}$ denotes the trace norm of a matrix $A\in\cL(\cH)$. Moreover, the optimal POVM for which \eqnsref{eq:err_class}{eq:err_quant} coincide is given by $\{\Pi_+,\,\Pi_-\}$, where $\Pi_+$ ($\Pi_-$) is the projection operator onto a non-negative (negative) subspace of the matrix $\pi_\rho \,\rho- \pi_\sigma \,\sigma$ appearing in \eqnref{eq:err_quant}~\cite{Helstrom1976}.

For equally likely hypotheses, $\pi_\rho=\pi_\sigma=\frac{1}{2}$, \eqnref{eq:err_quant} yields a direct generalisation of \eqnref{eq:p_min_err}:
\begin{equation}
p^{\textrm{err}}_{\min}=\frac{1}{2}\left(1-T[\rho,\sigma]\right),
\label{eq:trace_distance}
\end{equation}
where $T[\rho,\sigma]$ is now the \emph{trace distance}---the quantum equivalent of the TV distance \eref{eq:TV}---that is defined between any two quantum states $\rho,\sigma\in \cL(\cH)$ as
\begin{equation}
T[\rho,\sigma]:=\frac{1}{2}\left\Vert\rho-\sigma\right\Vert _{1}.
\label{eq:trace_dist}
\end{equation}
In fact, the trace distance can be equivalently obtained from the TV distance \eref{eq:TV} by following the procedure \eref{eq:cl_dist_to_q_dist}, so that it should not be surprising that it consistently satisfies all (\emph{i-iii}) conditions of \secref{subsec:cl_dist_div} forming a `proper' distance. Still, it can also be interpreted as an example of a quantum $f$-divergence defined for a convex function $f(t)=\frac{1}{2}|1-t|$ in \eqnref{eq:q_f_div}.

\paragraph{Asymptotic scenario ($n\to\infty$)}
For more than a single copy, $n>1$, the two hypotheses formally become $\rho^{\otimes n}$ and $\sigma^{\otimes n}$, so that although the adequate classical expression \eref{eq:chernoff_bound} for $p_{n,\trm{min}}^\trm{err}$ remains valid, in order to completely minimise it, one must allow for the freedom to perform \emph{collective measurements} (POVMs) on all the system copies at hand, see \figref{fig:Collective_hypot}(b). Note that only a single POVM outcome $i$ can then be assumed to be available, despite $n$ being even arbitrary large. On the other hand, we may interpret such a situation just as a single-shot quantum scenario, and substitute $\rho\to\rho^{\otimes n}$ and $\sigma\to\sigma^{\otimes n}$ in \eqnref{eq:err_quant} to obtain its equivalent for arbitrary $n$ as
\begin{equation}
p_{n,\trm{min}}^\trm{err}=\frac{1}{2}\left(1-\left\Vert\pi_\rho \,\rho^{\otimes n}- \pi_\sigma \,\sigma^{\otimes n}\right\Vert_{1}\right)
\quad\leq\quad 
\xi[\rho,\sigma]^n
\label{eq:chernoff_q}	
\end{equation}
which, as indicated above, can be generally upper-bounded by the \emph{quantum Chernoff bound} that takes an exactly analogous form to the (classical) Chernoff bound \eref{eq:chernoff_bound}~\cite{Audenaert2007,Calsamiglia2008}:
\begin{equation}
\xi[\rho,\sigma]=\min_{0\leq\alpha\leq1}\xi_{\alpha}[\rho,\sigma]\quad\textrm{with}\quad\xi_{\alpha}[\varrho,\sigma]\;:=\mathrm{Tr}\!\left\{ \varrho^{\alpha}\sigma^{1-\alpha}\right\} ,\label{eq:quant_Chernoff}
\end{equation}
being again independent of priors $\pi_\rho$ and $\pi_\sigma$, and symmetric with respect to the interchange of, this time, the quantum states $\rho$ and $\sigma$. The \emph{quantum Chernoff coefficients}, $\xi_\alpha$, are similarly to their classical equivalents not guaranteed to be symmetric for all $0\le\alpha\le1$ except $\alpha=1/2$, at which $\xi_{1/2}[\rho,\sigma]:=\tr{\sqrt{\rho}\sqrt{\sigma}}$ defines the so-called \emph{(quantum) affinity}~\cite{Luo2004b}. 

However, let us strongly emphasise that one should \emph{not} interpret the quantum affinity $\xi_{1/2}[\rho,\sigma]$ as a quantum equivalent of the Bhattacharyya coefficient, i.e.~$F[p,q]=\xi_{1/2}[p,q]=\sum_i\sqrt{p_i q_i}$ defined in \secref{sec:cl_HT_symm}. Although one may naively define the \emph{(squared) quantum Hellinger distance} as $H_{1/2}[\rho,\sigma]:=2(1-\xi_{1/2}[\rho,\sigma])$ and prove it to consistently satisfy all the necessary (\emph{i}-\emph{iii}) conditions of \secref{subsec:cl_dist_div} requisite for a distance~\cite{Luo2004b}, it does \emph{not} constitute the quantum version of classical Hellinger distance, $H_{1/2}[p,q]=2(1-F[p,q])$, in the sense of the procedure \eref{eq:cl_dist_to_q_dist}. If one performs explicitly the maximisation of $H_{1/2}[p,q]$ over POVMs in \eqnref{eq:cl_dist_to_q_dist}, one arrives instead at the so-called \emph{Bures distance}~\cite{bures1969extension}, $B[\rho,\sigma]:=2(1-F[\rho,\sigma])$, where
\begin{equation}
F[\rho,\sigma]:=\tr{\sqrt{\sqrt{\sigma}\rho\sqrt{\sigma}}}=\left\Vert\sqrt{\rho}\sqrt{\sigma}\right\Vert_{1}
\label{eq:fidelity}
\end{equation}
is the celebrated \emph{(quantum) fidelity}%
\footnote{In \citeref{jozsa1994fidelity} the quantum fidelity was originally introduced by Jozsa as generalisation of a ``transition probability'' rather than a distance-like measure, what would lead to an extra square in the definition \eref{eq:fidelity}, i.e.~$F[\rho,\sigma]:=\left|\sqrt{\rho}\sqrt{\sigma}\right|_{1}^2$.}~\cite{Uhlmann1976, jozsa1994fidelity},
 which is generally greater than or equal to affinity, i.e.~$\xi_{1/2}[\rho,\sigma]\leq F[\rho,\sigma]$, with the equality being assured for commuting $\rho$ and $\sigma$~\cite{Audenaert2014}.

Nonetheless, focussing on general quantum Chernoff coefficients, $\xi_\alpha$ in \eqnref{eq:quant_Chernoff}, in the same manner their classical counterparts define Hellinger divergences, $H_\alpha[p,q]$ in \eqnref{eq:Hell_div}, one may define with their help a natural family of quantum $f$-divergences \eref{eq:q_f_div} called Tsallis relative entropies~\cite{furuichi2004fundamental,Abe2003} that we term here as \emph{Tsallis divergences} to avoid confusion, i.e.:
\begin{equation}
H_\alpha[\rho,\sigma]:=\frac{1}{1-\alpha}(1-\xi_{\alpha}[\rho,\sigma])=\frac{1}{1-\alpha}\left(1-\tr{\rho^\alpha\sigma^{1-\alpha}}\right),
\label{eq:Tsall_div}
\end{equation}
which constitute an example of $D_f[\rho,\sigma]$ with $f(t)=\frac{1-t^\alpha}{1-\alpha}$ in \eqnref{eq:q_f_div} being convex and satisfying $f(1)=0$ for an extended range of $0\le\alpha<1$ and $\alpha>1$~\cite{Virosztek2016}, as required.

Returning to \eqnref{eq:chernoff_q} and considering the \emph{asymptotic scenario} of $n\to\infty$, a collective POVM of the form depicted in \figref{fig:Collective_hypot}(b) is crucially guaranteed to exist~\cite{Audenaert2008}, such that the inequality \eqnref{eq:chernoff_q} is attainable in a similar sense to the classical case and \eqnref{eq:as_prop_err_sHT}, i.e.~with the average probability for the optimal strategy decreasing exponentially (up to sublinear terms) as:
\begin{equation}
p_{n,\trm{min}}^\trm{err}
\quad\underset{n\to\infty}{=}\quad 
\exp\!\left[-n\,C[\rho,\sigma]+o(n)\right],
\label{eq:as_prop_err_sHT_quant}
\end{equation}
where the exponent corresponds now to a natural generalisation of \eqnref{eq:chernoff_info}, i.e.~the \emph{quantum Chernoff information} defined as~\cite{Audenaert2007}:
\begin{equation}
C[\rho,\sigma]:=-\lim_{n\to\infty}\frac{1}{n}\;\ln p_{n,\trm{min}}^\trm{err}=-\ln\xi[\rho,\sigma].
\label{eq:q_chernoff_info}
\end{equation}
However, let us stress that although \eqnref{eq:q_chernoff_info} consitutes a single-copy formula, being determined by the quantum Chernoff bound \eref{eq:quant_Chernoff}, there is no guarantee that there exist a \emph{local} (single-copy) measurement $\Pi_i$ in \figref{fig:Collective_hypot}(a), such that the classical Chernoff bound \eref{eq:clCB} for corresponding PDs $p_i=\mathrm{Tr}(\rho \Pi_i)$ and $q_i=\mathrm{Tr}(\sigma\Pi_i)$ coincides with its quantum counterpart \eref{eq:quant_Chernoff}. We discuss such a restriction to local POVMs explicitly below, but rather in the context of trace distance \eref{eq:trace_dist}.

Finally, in order to establish all the relevant quantities in the quantum setting, we note that the family of R\'{e}nyi divergences \eref{eq:Renyi_div} also possesses their quantum generalisation, i.e.~\cite{Hiai2017}:
\begin{align}
D_\alpha[\rho||\sigma]&:=\frac{1}{\alpha-1}\ln\!\left[1+(\alpha-1)H_{\alpha}[\rho,\sigma]\right] \label{eq:qRenyi_div}\\
&=\frac{1}{\alpha-1}\ln\xi_{\alpha}[\rho,\sigma]=\frac{1}{\alpha-1}\ln\left(\tr{\rho^{\alpha}\sigma^{1-\alpha}}\right),\nonumber
\end{align}
termed, as expected, \emph{quantum R\'{e}nyi divergences}, which similarly to the classical case constitute a non-decreasing function of Tsallis divergences \eref{eq:Tsall_div} and, thus, can be shown to be monotonic under quantum maps for $0\le\alpha<1$ and $1<\alpha<2$~\cite{Hiai2017}.

\paragraph{Restriction to local POVMs}
An important property of the \emph{fidelity} \eref{eq:fidelity} distinguishing it amongst other quantum distance-like measures is that it can be saturated by using only \emph{local} measurements of the form depicted in \figref{fig:Collective_hypot}(a). In order to demonstrate this, let us first reproduce the proof (c.f.~\cite{Nielsen2000}) of $F[\rho,\sigma]$ constituting the quantum equivalent of the Bhattacharyya coefficient $F[p,q]=\sum_{i}\sqrt{p_{i}q_{i}}$, i.e.~$F[\rho,\sigma]=\min_{\{\Pi_{i}\}_i}\sum_{i}\sqrt{p_{i}q_{i}}$ with $p_{i}=\tr{\rho\Pi_{i}}$ and $q_{i}=\tr{\sigma\Pi_{i}}$, which confirms that following the recipe \eref{eq:cl_dist_to_q_dist} the (classical) Hellinger distance generalises to the (quantum) Bures distances. Using the polar decomposition $\sqrt{\sqrt{\rho}\sigma\sqrt{\rho}}=\sqrt{\rho}\sqrt{\sigma}U$, the expression \eref{eq:fidelity} for fidelity can always be rewritten as
\begin{align}
F[\rho,\sigma]&=\tr{\sqrt{\rho}\sqrt{\sigma}U}=\tr{\sqrt{\rho}\,\sum_{i}\Pi_{i}\,\sqrt{\sigma}U} \nonumber\\
&=\sum_{i}\tr{\sqrt{\rho}\sqrt{\Pi_{i}}\sqrt{\Pi_{i}}\sqrt{\sigma}U} \nonumber\\
&\leq\sum_{i}\sqrt{\tr{\rho\Pi_{i}}\tr{\sigma\Pi_{i}}} \label{eq:dupa}\\
&=\sum_{i}\sqrt{p_{i}q_{i}}=F[p,q], \nonumber
\end{align}
where \eqnref{eq:dupa} constitutes a Cauchy-Schwarz inequality $\tr{A B}\leq \sqrt{\tr{A^\dagger A}\tr{B^\dagger B}}$ with $A=\sqrt{\rho}\sqrt{\Pi_i}$ and $B=\sqrt{\Pi_i}\sqrt{\sigma}U$ that may be saturated whenever for each outcome $i$ there exists $\lambda_i\in \RR$ such that $\sqrt{\Pi_{i}}\sqrt{\rho}=\lambda_{i}\sqrt{\Pi_{i}}\sqrt{\sigma}U$. For any invertible $\rho$ (and non-invertible ones by continuity), the above polar decomposition equivalently reads $\sqrt{\sigma}U=\left(\sqrt{\rho}\right)^{-1}\sqrt{\sqrt{\rho}\sigma\sqrt{\rho}}$, so that the condition for a POVM to saturate the inequality \eref{eq:dupa} can be restated as:
\begin{equation}
\forall_i:~\exists_{\lambda_i\in\RR}\quad\text{s.t.}\quad\sqrt{\Pi_{i}}\left(1-\lambda_{i}M_{\rho,\sigma}\right)=0
\label{eq:sat_cond}
\end{equation}
with
\begin{equation}
M_{\rho,\sigma}=\left(\sqrt{\rho}\right)^{-1}\sqrt{\sqrt{\rho}\sigma\sqrt{\rho}}\left(\sqrt{\rho}\right)^{-1}.
\label{eq:matrixM}
\end{equation}
Crucially, there always exists a POVM, $\{\Pi_{i}=|i\rangle\langle i|\}_i$ with $|i\rangle$ being the eigenvectors of the matrix $M_{\rho,\sigma}$ in \eqnref{eq:matrixM}, which satisfies the condition \eref{eq:sat_cond} and, hence, constitutes the optimal measurement that when performed guarantees the Bhattacharyya coefficient to coincide with the fidelity \eref{eq:fidelity}. Considering now the scenario of \figref{fig:Collective_hypot} with multiple ($n>1$) copies of the system available, all the above arguments again apply after simply replacing $\rho\to\rho^{\otimes n}$ and $\sigma\to\sigma^{\otimes n}$. Moreover, it is not hard to prove that the matrix $M$ in \eqnref{eq:matrixM} fulfils then $M_{\rho^{\otimes n},\sigma^{\otimes n}}=M_{\rho,\sigma}^{\otimes n}$, so that an optimal \emph{local} POVM, as in \figref{fig:Collective_hypot}(a), must always exist with its elements simply corresponding to the tensor product of the single-copy solution, i.e.~$\Pi_{i_1,\dots i_n}^{(n)}=\otimes_{m=1}^{n}|i_{m}\rangle\langle i_{m}|$.

This is in stark contrast to the case of \emph{trace distance} \eref{eq:trace_dist} for which a \emph{collective} measurement depicted in \figref{fig:Collective_hypot}(b) over multiple copies of the state is generally required, so that the TV distance \eref{eq:TV} between the resulting outcome PDs attains the optimal quantum value. In order to demonstrate this, let us consider first the recipe \eref{eq:cl_dist_to_q_dist} in the single-copy scenario, which for the trace distance reads $T[\rho,\sigma]=\max_{\{\Pi_{i}\}_i}T[p,q]$ with $p_{i}=\tr{\rho\Pi_{i}}$ and $q_{i}=\tr{\sigma\Pi_{i}}$. However, as we deal now with a maximisation over POVMs (differently to the case of fidelity), we rather focus on the r.h.s.~and upper bound the corresponding TV distance as follows:
\begin{align}
\frac{1}{2}\max_{\{\Pi_{i}\}_i}\sum_{i}|p_{i}-q_{i}|
&=
\frac{1}{2}\max_{\{\Pi_{i}\}_i}\sum_{i}|\tr{\Pi_{i}(\rho-\sigma)}| \nonumber\\
&\leq\frac{1}{2}\max_{\{\Pi_{i}\}_i}\sum_{i}\tr{\Pi_{i}|\rho-\sigma|} \nonumber\\
&=\frac{1}{2}\tr{|\rho-\sigma|}=T[\rho,\sigma].
\end{align}
Hence, in order to saturate the above inequality and crucially attain the trace distance, a binary projective measurement is required that projects onto the non-negative and negative subspaces of the matrix $\rho-\sigma$, as in \secref{sec:sHT_quantum}. Turning to the multiple copy ($n>1$) scenario and substituting for $\rho\to\rho^{\otimes n}$ and $\sigma\to\sigma^{\otimes n}$ above, it becomes clear that it is now the $\rho^{\otimes n}-\sigma^{\otimes n}$ matrix that must be decomposed into its (non-)negative subspaces to construct the optimal POVM, which consequently must be in principle collective as in \figref{fig:Collective_hypot}(b), and act on all the system $n$ copies.

\subsubsection{Asymmetric Quantum Hypothesis Testing}
\label{sec:aHT_quantum}
Adopting the task of aHT introduced in \secref{sec:cl_HT_asymm} to the quantum setting, one must still perform minimisation of the respective one-type error probability---taken again to be the type-II error $P(p^{(n)}|q^{(n)})$ in \eqnref{eq:error_prob_sHT}---where the integer $n$, however, denotes now the number of independent copies of the quantum system available rather than independent rounds of the protocol. In particular, while keeping $P(q^{(n)}|p^{(n)})\leq \epsilon$ for some $\epsilon>0$, $P(p^{(n)}|q^{(n)})$ must be minimised over all possible POVMs $\{\Pi_i\}_i$, in particular, \emph{collective} ones depicted in \figref{fig:Collective_hypot}(b), which may yield only a single outcome $i$ distributed then according to either $p^{(n)}_i=\textrm{Tr}\{\rho^{\otimes n}\Pi_i\}$ or $q^{(n)}_i=\textrm{Tr}\{\sigma^{\otimes n} \Pi_i\}$. Nonetheless, in the asymptotic regime of infinitely many system copies ($n\to\infty$), one may prove the number $n$ to be treatable at the same grounds as the repetition number in the classical case of aHT, so that $P(p^{(n)}|q^{(n)})$ minimised over all POVMs decays exponentially in a similar manner to \eqnref{eq:assym_entr}, i.e.~\cite{Hiai1991,Ogawa2000}:
\begin{equation}
P_{\trm{min}}\!\left(\rho^{\otimes n}|\sigma^{\otimes n}\right)
\quad\underset{n\to\infty}{=}\quad 
\exp\!\left[-n\,D[\rho||\sigma]+o(n)\right],
\label{eq:assym_quant-1}
\end{equation}
where $D[\rho\|\sigma]$ above is now the \emph{quantum relative entropy} of the (single-copy) state $\rho$ with respect to the state $\sigma$, taking the form originally introduced by Umegaki~\cite{umegaki1962conditional}:
\begin{align}
D[\rho||\sigma]&\equiv\lim_{\alpha\to1^-}D_\alpha[\rho||\sigma]\;\equiv\lim_{\alpha\to1^-}H_\alpha[\rho||\sigma]\nonumber\\
&:=\tr{\rho\left(\ln\rho-\ln\sigma\right)}.
\label{eq:q_rel_entr}
\end{align}
We have noted above that the quantum relative entropy $D[\rho||\sigma]$---similarly to its classical counterpart (Kullback-Leibler divergence) $D[p||q]$ in \eqnref{eq:rel_entr} that is interpretable as a (one-sided) limit of Hellinger \eref{eq:Hell_div} and R\'{e}nyi \eref{eq:Renyi_div} divergences---can be analogously defined with help of the quantum Tsallis \eref{eq:Tsall_div} and R\'{e}nyi \eref{eq:qRenyi_div} divergences~\cite{furuichi2004fundamental,Hiai2017}, while constituting thus naturally a quantum $f$-divergence \eref{eq:q_f_div} with $f(t)=-\ln t$~\cite{Hiai2017}. 

Moreover, as for the (classical) relative entropy $D[p||q]$ that is finite only if $\sup p\subseteq\sup q$ in \eqnref{eq:rel_entr}, the quantum relative entropy $D[\rho||\sigma]$ becomes divergent whenever the support of $\rho$ is not contained in the support of $\sigma$ in the matrix sense, i.e.~$\sup \rho\nsubseteq\sup \sigma$. For instance, if $\rho$ is a pure state then $D[\rho||\sigma]=\infty$ iff $\rho\neq\sigma$. This should be expected, since had one measured then in the eigenbasis of $\rho$ and obtained a result prohibited under the assumption that the correct state is $\sigma$, one would become sure about the correctness of state $\rho$---an event that consistently can never occur, as then $P_{\trm{min}}\!\left(\rho^{\otimes n}|\sigma^{\otimes n}\right)=0$ in \eqnref{eq:assym_quant-1} due to $D[\rho||\sigma]=\infty$.

\subsection{Information geometry of quantum states}
As indicated already in \figref{fig:stat_man}, the geometrical approach described in \secref{subsec:Information-geometry} for the classical theory can be straightforwardly
generalised to the quantum picture~\cite{Amari2000,Hayashi2006}. Each state $\rho\in\cL(\cH)$, similarly to each PD $p\in\cM$, can be treated just as a point on a Riemannian manifold $\cMq$ known as the \emph{quantum statistical manifold}. Furthermore, analogously to the case of (classical) statistical manifold $\cM$, one may introduce a local coordinate system on $\cMq$, so that all the contained quantum states, $\rho_{\mathbb{\paramV}}\in\cMq$, become parametrised by a vector $\paramV\in\RR^{d}$ of dimension $d=\textrm{dim}(\cMq)$. Similarly, all other definitions from \secref{subsec:Information-geometry}, since they describe just differential-geometric concepts, transfer directly to encompass the geometry of quantum states. We will therefore just briefly summarise basic notions of quantum information geometry, leaving the emerging features that distinguish it from the classical case for next sections.

\subsubsection{Divergence-induced quantum metrics}
As in the classical setting of \secref{sec:div_metr_class}, any quantum divergence $D$ that satisfies the condition (\emph{ii}) of \secref{subsec:cl_dist_div} and is smooth in both of its arguments, must induce a metric on the quantum statistical manifold $\cMq$. In particular, for any two vectors $V,\,W \in \tT_\rho \cM$ lying in the tangent space at a given point $\rho\in\cMq$ one can define the $D$-induced metric $\g_D$ and its matrix representation $\gM_D$ analogously to \eqnsref{eq:div_ind_metric0}{eq:div_ind_metric1} as:
\begin{align}
\g_D(\rho)\left[V,W\right]&=V^T \gM_D(\rho)W \label{eq:quantum_metric}\\
&=-\sum_{ij}\left.\frac{\partial^{2}}{\partial t\partial s}D\!\left[\rho+tV_{i},\rho+sW_{j}\right]\right|_{t=s=0},\nonumber
\end{align}
which measures now susceptibility of the respective divergence $D$ to small changes of the quantum state $\rho\in\cMq$. Similarly to \eqnref{eq:div_ind_metric_phi}, assuming some local coordinate system $\paramV$ induced on $\cMq$, a change of the divergence under small deviation $\dparamV$ of the vector $\paramV$ can be written as
\begin{equation}
D\!\left[\rho_{\paramV},\rho_{\paramV+\dparamV}\right]
\approx
\frac{1}{2}\dparamV^{T}\;\gMp_D(\rho_{\paramV})\!\left[\nabla\rho_{\paramV},\nabla\rho_{\paramV}\right]\;\dparamV.
\label{eq:div_metric_q}
\end{equation}
Note, that since $D\!\left[\rho_{\paramV},\rho_{\paramV}\right]=0$ is the minimal value of a divergence, the first-order term $\Theta(\dparamV)$ must be absent from the above expression and one obtains the same metric $\gMp_D(\rho_\paramV)$ as a matrix the in $\paramV$-coordinate system regardless of which argument is perturbed, i.e.~$D\!\left[\rho_{\paramV},\rho_{\paramV+\dparamV}\right]=D\!\left[\rho_{\paramV+\dparamV},\rho_{\paramV}\right] + O(\delta\varphi^2)$, even though the divergence itself may not be symmetric. However, for the same reason $\sup(p_{\paramV+\boldsymbol{\delta \param}})\nsubseteq\sup(p_\paramV)$ implies the metric $\gMp_D(p_\paramV)$ in  \eqnref{eq:metric_coordinates} to be ill-defined in the classical case, whenever $\sup(\rho_{\paramV+\boldsymbol{\delta \param}})\nsubseteq\sup(\rho_\paramV)$, i.e.~the state $\rho_{\paramV+\boldsymbol{\delta \param}}$ changes its rank exactly at $\boldsymbol{\delta \param}=0$, the metric $\g_D(\rho_\paramV)$ in \eqnref{eq:quantum_metric} does not exist and the corresponding $\paramV$-based expansions \eref{eq:div_metric_q} of divergences (e.g.~of the Bures distance~\cite{Siafranek2017, Seveso2019}) cannot be performed.

Importantly, as was the case in the classical setting, monotonic divergences induce monotonic metrics. The argument is similar as before with a general stochastic map $\m{S}$ replaced now by a general quantum map $\Lambda$. From monotonicity of a quantum divergence it follows that $D[\Lambda[\rho_\paramV],\Lambda[\rho_\paramV+\nabla\rho_{\paramV}\delta\paramV]]\leq D[\rho_\paramV,\rho_\paramV+\nabla\rho_{\paramV}\delta\paramV]$ and, since $\Lambda$ is linear and does not depend on $\paramV$, by \eqnref{eq:div_metric_q} one obtains $\gMp_D(\Lambda[\rho_\paramV])[\nabla \Lambda[\rho_\paramV],\nabla \Lambda[\rho_\paramV]]\leq \gMp_D(\rho_\paramV)[\nabla \rho_\paramV,\nabla \rho_\paramV]$, as required.

A particularly useful class of metrics are the ones induced by the quantum $f$-divergences \eref{eq:q_f_div}. These may be obtained by expanding any given quantum $f$-divergence for neighbouring states $\rho_\paramV$ and $\rho_{\paramV+\dparamV}$ (with $\sup(p_{\paramV+\boldsymbol{\delta \param}})\subseteq\sup(p_\paramV)$) in the $\paramV$-coordinate system as in \eqnref{eq:div_metric_q}, so that
\begin{equation}
D_{f}\!\left[\rho_{\paramV},\rho_{\paramV+\dparamV}\right]
\approx
\frac{1}{2}\dparamV^{T}\,\gMp_{D_f}\!(\rho_{\paramV})\,\dparamV,
\end{equation}
where
\begin{equation}
\gMp_{D_f}(\rho_{\paramV})=\ddot{f}(1)\,\QFM{f}(\rho_{\paramV})
\label{eq:divergence_f_QFI}
\end{equation}
and the (\emph{quantum}) \emph{$f$-metric} above, $\QFM{f}(\rho_{\paramV})$, should be interpreted as the quantum generalisation of the Fisher metric \eref{eq:Fisher_metric}. We have omitted the (tangent-vector) arguments $\left[\nabla\rho_{\paramV},\nabla\rho_{\paramV}\right]$ as before, but in contrast to \eqnref{eq:Fisher_metric} we have explicitly kept the dependence on the convex function $f$ specifying a given quantum $f$-divergences \eref{eq:q_f_div} for reasons that will soon become clear.

Still, without defining explicitly $\QFM{f}$ in \eqnref{eq:divergence_f_QFI} for a given $D_f$, any $f$-metric \eref{eq:divergence_f_QFI} must inherit properties of a general divergence-induced metric defined in \eqnref{eq:div_metric_q}. In particular, $\QFM{f}$ similarly to $\gMp_D$ must be \emph{additive} on tensor products of quantum states, i.e.~$\QFM{f}\!\left(\rho_\paramV\otimes\sigma_\paramV\right)=\QFM{f}\!\left(\rho_\paramV\right)+\QFM{f}\!\left(\sigma_\paramV\right)$, which is a reminiscent of the additivity property of metrics on Cartesian products of statistical manifolds discussed in \secref{subsec:cl_metrics}. Moreover, thanks to the \emph{monotonicity} under quantum maps inherited by $\gMp_D$ in \eqnref{eq:div_metric_q} from any monotonic divergence $D$, another feature that transfers straightforwardly from the classical setting and can be proven analogously to \secref{subsec:FI_metric}, is the \emph{convexity} property of any $f$-metric \eref{eq:divergence_f_QFI}. In particular, for any states $\rho_\paramV,\sigma_\paramV\in\cMq$ and any $\lambda\in[0,1]$ it generally holds that $\QFM{f}(\lambda\rho_\paramV+(1-\lambda)\sigma_\paramV)\leq\lambda\QFM{f}(\rho_\paramV)+(1-\lambda)\QFM{f}(\sigma_\paramV)$.

\subsection{Plethora of metrics monotonic under quantum maps}
\label{sec:plethora}
One of the profound differences between the geometrical structure of the classical $\cM$ and quantum $\cMq$ statistical manifolds depicted in \figref{fig:stat_man} is the fact that $\cMq$ in contrast to $\cM$ allow for many inequivalent (monotonic) metrics defined based on quantum $f$-divergences \eref{eq:q_f_div}. Considering a $\cMq$ parametrised by a particular coordinate system $\paramV$, one can explicitly write the $f$-metric introduced in \eqnref{eq:divergence_f_QFI} at a given $\rho_{\paramV}\in\cMq$ as~\cite{Petz2002,Lesniewski1999}:
\begin{align}
&\QFM{f}(\rho_{\paramV})_{ij}=\sum_{n}\frac{\partial_{i}p_{n}\partial_{j}p_{n}}{p_{n}}+ \label{eq:fQFI_1}\\
&\qquad+\frac{1}{\ddot{f}(1)}\sum_{m,n}\left[p_{m}f\!\left(\frac{p_{n}}{p_{m}}\right)+p_{n}f\!\left(\frac{p_{m}}{p_{n}}\right)\right]\langle n|\partial_{i}m\rangle\langle\partial_{j}m|n\rangle, \nonumber
\end{align}
where $p_{n}$ and $|n\rangle$ are the $\paramV$-dependent eigenvalues and eigenvectors of the quantum state $\rho_{\paramV}=\sum_n p_n(\paramV)\proj{n(\paramV)}$, and by $\partial_{i}$ we denote the derivative with respect to the $i$th component of $\paramV$. Note that, as expected from the `quantum to classical correspondence' applying for diagonal density matrices, the first term above corresponds to nothing but the (classical) Fisher metric, $\FM(p_\paramV)$ in \eqnref{eq:Fisher_metric}, computed for the $\paramV$-parametrised PD defined by the eigenvalues of $\rho_{\paramV}$.

It is convenient to also rewrite the expression for the $f$-metric \eref{eq:fQFI_1} by reparametrising the convex function $f:\RR_+\to\RR_+$ associated with the quantum $f$-divergence \eref{eq:q_f_div} by another function $g:\RR_+\to\RR_+$ such that for all $t\in\RR_+$ it fulfils: 
\begin{equation}
\label{eq:f_g}
\frac{1}{g(t)}=\frac{f(t)+tf(1/t)}{(t-1)^{2}}.
\end{equation}
Moreover, one may set $\ddot{f}(1)=1$ in \eqnsref{eq:divergence_f_QFI}{eq:fQFI_1} for simplicity, which can always be assured by just rescaling the $D_f$ being considered in \eqnref{eq:divergence_f_QFI}. As a result, another classification of (quantum) metrics parametrised by functions $g$ is obtained:
\begin{equation}
\QFM{g}(\rho_{\paramV})_{ij}=\sum_{n}\frac{\partial_{i}p_{n}\partial_{j}p_{n}}{p_{n}}+\sum_{m,n}\frac{\left(p_{n}-p_{m}\right)^{2}}{p_{m}\,g\!\left(\frac{p_{n}}{p_{m}}\right)}\langle n|\partial_{i}m\rangle\langle\partial_{j}m|n\rangle,
\label{eq:g_metric}
\end{equation}
which we refer to as the \emph{$g$-metrics} to distinguish them from the $f$-metrics specified in \eqnref{eq:fQFI_1}.

The existence of inequivalent monotonic metrics on the quantum statistical manifold $\cMq$ is the essence of the \emph{quantum Chentsov theorem}~\cite{Morozova1991,Petz1996}. The theorem states that every Riemannian metric $\g$ on $\cMq$ that is \emph{monotonic} under the action of quantum maps must be (up to a multiplicative factor) proportional to one of the $g$-metrics specified in \eqnref{eq:g_metric} with a function $g$ that constitutes a \emph{standard operator monotone}, i.e.~a function $g:\RR_+\to\RR_+$ that satisfies $\forall_{t\in\RR_+}\!:\,g(t)=t\,g(1/t)$ and $g(1)=1$, and when naturally generalised to positive semidefinite matrices (see \secref{subsec:quant_f-div}) fulfils $g(A)\leq g(B)$ for any pair of positive matrices such that $A\leq B$. Interestingly, all such $g$-functions are identified by ones that for all $t\in\RR_+$ satisfy the inequality~\cite{Petz1996,Lesniewski1999}:
\begin{equation}
\frac{2t}{t+1}\leq g(t)\leq\frac{1+t}{2}.
\label{eq:g_inequality}
\end{equation}

Crucially, as any $f$-metric \eref{eq:fQFI_1} is guaranteed to be monotonic thanks to being induced by a (monotonic) quantum $f$-divergence \eref{eq:q_f_div}, it must always be expressible as a $g$-metric \eref{eq:g_metric} with a $g$-function satisfying the inequality \eref{eq:g_inequality}. This is consistently true, as for any convex function $f$ (with $f(1)=0$) specifying a $D_f$ in \eqnref{eq:q_f_div}, the corresponding function $g$ defined through the relation \eref{eq:f_g} constitutes always up to the rescaling factor $\ddot{f}(1)$ (set here to unity for convenience) a standard operator monotone~\cite{Lesniewski1999}.

In conclusion, in stark contrast to the classical case described in \secref{subsec:FI_metric}, quantum $f$-divergences \eref{eq:q_f_div} induce metrics that cannot be trivially related, while explicitly depending on a particular choice of the function $f$ assumed in \eqnref{eq:q_f_div}~\cite{Petz2002,Petz1996,Lesniewski1999}. More generally, depending on the (monotonic) quantum divergence $D$ used to distinguish states in a quantum statistical manifold $\cMq$, a different notion of the inner product in the tangent space, $\tT_\rho\cMq$ for any $\rho\in\cMq$ in \figref{fig:stat_man}, is obtained, as the metric $\g_D$ being induced strongly varies with the choice of $D$ in the quantum setting. In what follows, we discuss in more detail several important metrics that may so arise, while resorting to the classification based on the $g$-metrics \eref{eq:g_metric}.

\begin{figure}[t!]
\centering
\includegraphics[width=\columnwidth]{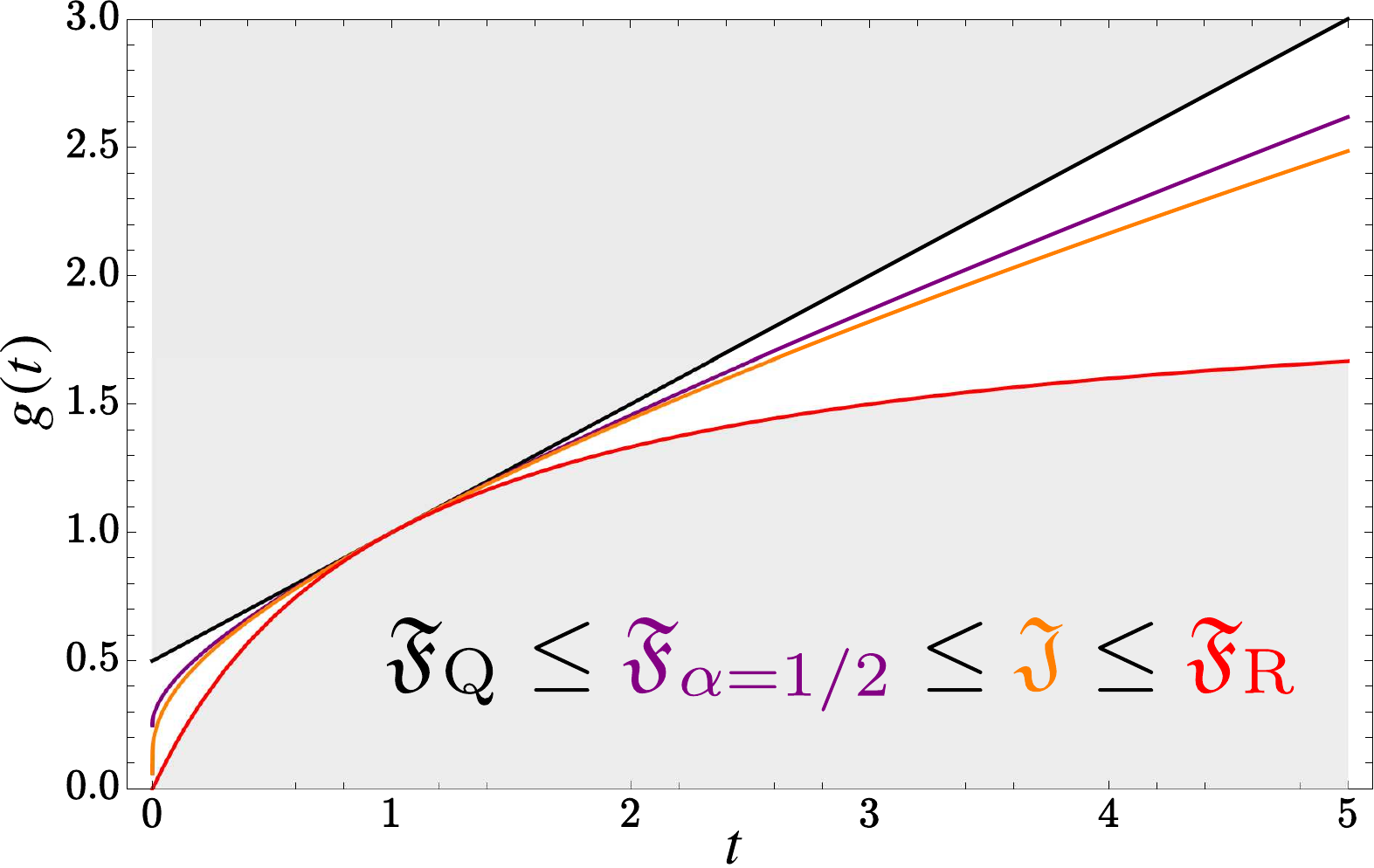}
\caption{{\bf Standard operator monotone $g$-functions and the monotonic $g$-metrics}, $\QFM{g}$ in \eqnref{eq:g_metric}, they generate. The inequalities \eref{eq:g_inequality} forbid any $g(t)$ to enter the \emph{shaded regions}, while the ordering of $g$-functions ($g\ge g'$) leads to a reversed order of their corresponding $g$-metrics ($\QFM{g}\le\QFM{g'}$). As a result, the maximal $g(t)=\frac{1+t}{2}$ (\emph{black}) yields the QFI metric $\QFIM$ that is minimal in the hierarchy \eref{eq:qmetric_hierachy}, whereas the minimal $g(t)=\frac{2t}{1+t}$ (\emph{red}) yields the maximal RLD metric $\QFMR$. We also mark $g(t)=\frac{1}{4}(1+\sqrt{t})^2$ (\emph{purple}) that generates the WYD metric with $\alpha=1/2$, $\QFM{\alpha=1/2}$ in \eqnref{eq:WYD_metric}, as well as $g(t)=\frac{t-1}{\ln t}$ (\emph{orange}) that yields the KM metric, $\KM$ in \eqnref{eq:KM_metric}.
}
\label{fig:g_function}
\end{figure}

\subsubsection{Quantum Fisher Information and Right Logarithmic Derivative metrics}
By choosing a $g$-metric \eref{eq:g_metric} with the maximal possible function $g$ fulfilling the inequality \eref{eq:g_inequality}, one arrives at the so-called \emph{quantum Fisher information} (QFI) metric, $\QFIM:=\QFM{g}$ with $g(t)=\frac{1+t}{2}$, which has an important practical interpretation in quantum information theory. It can be easily verified with \eqnref{eq:g_metric} that the QFI-metric can be equivalently defined as~\cite{Braunstein1994,Helstrom1976,Holevo2001}:
\begin{equation}
\QFIM(\rho_{\paramV})_{ij}=\frac{1}{2}\tr{\rho_{\paramV}\left(\mathcal{L}_{\param_{i}}\mathcal{L}_{\param_{j}}+\mathcal{L}_{\param_{j}}\mathcal{L}_{\param_{i}}\right)},
\label{eq:QFI_metric}
\end{equation}
where $\mathcal{L}_{\param_{i}}$ are the so-called \emph{symmetric logarithmic derivative} (SLD) operators, each defined implicitly as a solution to the equation $\frac{\partial\rho_{\paramV}}{\partial\param_{i}}=\frac{1}{2}\left(\rho_{\paramV}\mathcal{L}_{\param_{i}}+\mathcal{L}_{\param_{i}}\rho_{\paramV}\right)$ with $\frac{\partial}{\partial\param_{i}}$ denoting an entry-wise derivative of the state $\rho_\paramV$ with respect to the $i$th $\paramV$-coordinate. From the operational perspective of quantum metrology~\cite{BrandorffNielsen2000, Lee2002, Giovannetti2006, Giovannetti2011}, which we refer to later in \secref{subsec:Quantum-metrology}, by the virtue of so-called quantum Cram\'{e}r-Rao inequality the QFI metric \eref{eq:QFI_metric} can be used to lower bound the minimal mean-squared-error when treating $\paramV$ as a vector of parameters to be estimated. Lastly, let us note that
the \emph{geodesic distance} (see \secref{subsec:cl_metrics}) defined by the QFI-metric takes a compact form---it corresponds to the \emph{Bures angle} $B_\trm{A}[\rho,\sigma]:=\arccos(F[\rho,\sigma])$~\cite{Sommers2003}, which should be understood as a modification of the Bures distance $B[\rho,\sigma]$ introduced in \secref{sec:sHT_quantum} to a surface of a sphere.

On the other hand, by choosing a $g$-metric \eref{eq:g_metric} with the minimal possible function $g$ fulfilling the inequality \eref{eq:g_inequality}, one arrives at the \emph{right logarithmic derivative} (RLD) metric, $\QFMR:=\QFM{g}$ with $g(t)=\frac{2t}{t+1}$, which can be defined similarly to \eqnref{eq:QFI_metric} after replacing the SLD operators, $\mathcal{L}_{\param_{i}}$, with the so-called RLD operators $\mathcal{R}_{\param_{i}}$ defined implicitly by the equation $\frac{\partial\rho_{\paramV}}{\partial\param_{i}}=\rho_{\paramV}\mathcal{R}_{\param_{i}}$~\cite{Helstrom1976,Holevo2001}. 

It is not hard to verify that for any two standard operator monotone functions $g_1(t)\leq g_2(t)$ the corresponding $g$-metrics \eref{eq:g_metric} fulfil $\QFM{g_1}\geq\QFM{g_2}$ in the matrix sense. Hence, the maximal and minimal $g$-functions defined by the inequality \eref{eq:g_inequality}, yield in fact the minimal and maximal $g$-metrics, respectively, i.e.~for any $\rho_\paramV\in\cMq$ it generally holds that:
\begin{equation}
\QFIM(\rho_{\paramV})\leq\QFM{g}(\rho_{\paramV})\leq\QFMR(\rho_{\paramV}),
\label{eq:qmetric_hierachy}
\end{equation}
where the QFI and RLD metrics thus define the extremal (minimal and maximal) cases of the so-constructed hierarchy of (monotonic) $g$-metrics \eref{eq:g_metric} satisfying the inequality \eref{eq:g_inequality}~\cite{Petz1996}. In \figref{fig:g_function} we demonstrate explicitly how the hierarchy \eref{eq:qmetric_hierachy} arises, by plotting the minimal and maximal $g$-functions allowed by \eqnref{eq:g_inequality}, as well as ones generating the Wigner-Yanase-Dyson and Kubo-Mori metrics that we will now discuss. In particular, as $g_1\ge g_2$ implies $\QFM{{g_1}}\le\QFM{{g_2}}$, thanks to the functional inequalities between the corresponding $g$-functions evident from the plot, one may directly deduce the matrix inequalities between the $g$-metrics they generate.

\subsubsection{Wigner-Yanase-Dyson metrics}
\label{sub:WYD}
Let us consider the Tsallis divergences \eref{eq:Tsall_div} that constitute an example of quantum $f$-divergences with $f(t)=\frac{1-t^\alpha}{1-\alpha}$ in \eqnref{eq:q_f_div} satisfying $\ddot{f}_\alpha(1)=\alpha$. Hence, by adequately defining a \emph{rescaled} function $\tilde{f}_\alpha(t)=\frac{1-t^\alpha}{\alpha(1-\alpha)}$ with $\ddot{\tilde{f}}_\alpha(1)=1$, we can directly identify the metric induced by every $D_{\tilde{f}_\alpha}$ with help of \eqnref{eq:f_g} as a $\alpha$-parametrised class of $g$-metrics \eref{eq:g_metric} with $g_{\alpha}(t)=\alpha(1-\alpha)\frac{(t-1)^{2}}{(t^{\alpha}-1)(t^{1-\alpha}-1)}$.  The functions $g_{\alpha}(t)$ satisfy the inequality \eref{eq:g_inequality} for any $t\in\RR_+$ and, hence, constitute standard operator monotones for $-1\leq\alpha\leq 2$, for which they define the so-called \emph{(generalised) Wigner-Yanase-Dyson} (WYD) metrics~\cite{Manko2017}:
\begin{align}
&\QFM{\alpha}(\rho_{\paramV})_{ij}
=\sum_{n}\frac{\partial_{i}p_{n}\partial_{j}p_{n}}{p_{n}} + \label{eq:WYD_metric}\\
&\qquad+\frac{2}{\alpha(1-\alpha)}\sum_{n,m}p_{m}^{1-\alpha}\left(p_{m}^{\alpha}-p_{n}^{\alpha}\right)\langle n|\partial_{i}m\rangle\langle\partial_{j}m|n\rangle,\nonumber
\end{align}
which for any $\rho_\paramV\in\cMq$ generally fulfil $\QFM{\alpha=1/2}\le\QFM{\alpha}\le\QFM{\alpha=-1}=\QFM{\alpha=2}=\QFMR$ being constrained by the (monotonic) metric hierarchy \eref{eq:qmetric_hierachy}. The minimal WYD metric $\QFM{\alpha=1/2}$ constitutes a special $g$-metric with $g(t)=\frac{1}{4}(1+\sqrt{t})^2$, for which a geodesic distance can be explicitly computed as for the QFI metric \eref{eq:QFI_metric}. It takes a similar form of $D_\trm{WYD}[\rho,\sigma]:=\arccos(\xi_{1/2}[\rho,\sigma])$, where the ``spherical angle'' is now dictated by the affinity $\xi_{1/2}[\rho,\sigma]$ rather than the fidelity $F[\rho,\sigma]$ between quantum states $\rho,\sigma\in\cMq$~\cite{Gibilisco2003wigner}.

In the special case of states $\bar{\rho}_\paramV=\sum_n p_n\proj{n(\paramV)}$, whose eigenvalues $p_n$ are independent of $\paramV$ so that the first term (the classical Fisher-metric \eref{eq:Fisher_metric} contribution) vanishes in \eqnref{eq:WYD_metric}, one can assign to each WYD metric $\QFM{\alpha}$ a quantity called the \emph{generalised Wigner-Yanase-Dyson} (WYD) \emph{divergence}%
\footnote{For consistency, we adopt the usual ``misleading'' name for $\WYD_{\alpha}$ as the WYD \emph{divergence}, however, let us emphasise that it is unrelated to the canonical meaning of \emph{quantum divergences} that measure distances between quantum states, see \secref{subsec:quant_dist_meas}.} that still forms a matrix in the $\paramV$-coordinate system, i.e.~\cite{Wigner1963,Lieb1973}:
\begin{equation}
\WYD_{\alpha}(\bar{\rho}_{\paramV}):=\frac{\alpha(1-\alpha)}{2}\,\QFM{\alpha}(\bar{\rho}_{\paramV}),
\label{eq:WYD_div}
\end{equation}
but in contrast to $\QFM{\alpha}$ is maximal for $\alpha=\frac{1}{2}$, i.e.~$\WYD_{\alpha}(\bar{\rho}_\paramV)\le\WYD_{1/2}(\bar{\rho}_\paramV)$ for any $\bar{\rho}_\paramV\in\cMq$. The maximal $\WYD_{1/2}(\bar{\rho}_\paramV)$ is commonly referred to as the \emph{Wigner-Yanase} (WY) \emph{skew divergence}~\cite{Wigner1963}.

Although the WYD divergences $\WYD_{\alpha}$ in \eqnref{eq:WYD_div} do not formally constitute metrics---in contrast to the WYD metrics $\QFM{\alpha}$ in \eqnref{eq:WYD_metric}---the WY skew-divergence $\WYD_{1/2}$ turns out to be particularly useful in quantum information theory as a measure of coherence~\cite{Marvian2014,Girolami2014} and correlations~\cite{Luo2012} of quantum states. Moreover, it can be conveniently used to lower- and upper-bound the QFI metric $\QFIM$ in \eqnref{eq:QFI_metric} for any (eigenvalue-invariant) $\bar{\rho}_\paramV\in\cMq$ through inequalities~\cite{Luo2004}:
\begin{equation}
4\,\WYD_{1/2}(\bar{\rho}_{\paramV}) \;\leq\; \QFIM(\bar{\rho}_{\paramV}) \;\leq\; 8\,\WYD_{1/2}(\bar{\rho}_{\paramV}).
\label{eq:QFI_WYD-1}
\end{equation}
In order prove the left inequality above, it is convenient to explicitly write
\begin{equation}
4\,\WYD_{1/2}(\bar{\rho}_{\paramV})_{ij}=4\sum_{n,m}\sqrt{p_{m}}\left(\sqrt{p_{m}}-\sqrt{p_{n}}\right)\langle n|\partial_{i}m\rangle\langle\partial_{j}m|n\rangle.
\end{equation}
and interpret it as a (non-monotonic) $g$-metric at $\bar{\rho}_\paramV$ with a function $g(t)=\frac{1}{2}\left(\sqrt{t}+1\right)^{2}$ in \eqnref{eq:g_metric}. However, as the $g(t)$ is then larger for any $t\in\RR_+$ than the maximal behaviour allowed by the inequality \eref{eq:g_inequality}---and, hence, does not constitute a standard operator monotone---its resulting $g$-metric must be smaller than the one associated to the maximal $g(t)=\frac{1+t}{2}$, i.e.~the QFI metric $\QFIM(\bar{\rho}_{\paramV})$ defined in \eqnref{eq:QFI_metric}. On the other hand, the rightmost inequality in \eqnref{eq:QFI_WYD-1} directly follows from \eqnref{eq:qmetric_hierachy}, which generally implies $\QFIM \leq \QFM{\alpha=1/2}$ and, hence, $\QFIM(\bar{\rho}_{\paramV}) \leq 8\,\WYD_{1/2}(\bar{\rho}_{\paramV})$  
for any $\bar{\rho}_\paramV\in\cMq$ after substituting for the WY skew divergence according to \eqnref{eq:WYD_div}.

\subsubsection{Kubo-Mori metric}
The last monotonic metric we would like to mention is the \emph{Kubo-Mori} (KM) metric~\cite{Kubo1980means,Petz1996}, see also~\cite{Koenig2013}, that constitutes a $g$-metric \eref{eq:g_metric} with a (standard operator monotone) function $g(t)=\frac{t-1}{\ln t}$ and for a general $\rho_\paramV\in\cMq$ reads
\begin{equation}
\KM(\rho_{\paramV})_{ij}=\sum_{n}\frac{\partial_{i}p_{n}\partial_{j}p_{n}}{p_{n}}
+2\!\sum_{n,m}(p_{n}-p_{m})\ln p_{n}\,\langle n|\partial_{i}m\rangle\langle\partial_{j}m|n\rangle.
\label{eq:KM_metric}
\end{equation}
Note that it can be equivalently interpreted as the WYD metric after taking the limit of $\alpha\to0\lor1$ in \eqnref{eq:WYD_metric}, i.e.~$\KM=\QFM{\alpha=0}=\QFM{\alpha=1}$, while also naturally falling into the mid-range of the metric hierarchy \eref{eq:qmetric_hierachy}. From the quantum information perspective, the KM metric covers the effects of output superadditivity in classical communication over quantum channels~\cite{Czajkowski2016}, and naturally appears also in the contexts of thermodynamics~\cite{Crooks2007} and renormalisation theory~\cite{Beny2015}.

\subsection{Selected scalar metrics for pure states and the unitary parametrisation}
\label{sec:unitary}
In general, the $g$-metrics take a complicated form \eref{eq:g_metric} which is hard to deal with, especially when one's aim is to further optimise the metric over the quantum states. However, for families of states parametrised by a \emph{single} parameter $\param$, the $g$-metrics $\QFM{g}$ become (non-negative) \emph{scalars}, which we denote by $\QFIg{g}$ and call \emph{$g$-informations} for short. Moreover, if the parameter is encoded \emph{unitarily} in a linear fashion, i.e.~by a unitary quantum map $\varrho_{\param}=U_{\param}\varrho U_{\param}^{\dagger}$ with $U_{\param}=\ee^{-\ii\param \hat H}$ for a given fixed Hamiltonian $\hat H$, the expression \eref{eq:g_metric} for the $g$-information simplifies significantly to 
\begin{equation}
\label{eq:gmetric_unitary}
\QFIg{g}\!\left(\varrho, \hat H\right)=\sum_{n,m}\frac{\left(p_{n}-p_{m}\right){}^{2}}{p_{m}\,g\!\left(\frac{p_{n}}{p_{m}}\right)}|H_{nm}|^{2},
\end{equation}
where $\varrho=\sum_n p_n \proj{n}$ and $H_{nm}=\langle n|\hat H |m\rangle$. In \eqnref{eq:gmetric_unitary} we have explicitly written that the $g$-information for the unitary encoding is fully defined by the (unperturbed) state $\varrho$ and the Hamiltonian $\hat H$, while $\QFIg{g}$ is independent of the actual value of the parameter $\param$. 

On one hand, if the state is pure $\varrho=\proj{\psi}$, \eqnref{eq:gmetric_unitary} simplifies even further to
\begin{equation}
\label{eq:QFI_g_pure}
\QFIg{g}\!\left(\ket{\psi}, \hat H\right)=\frac{2}{g(0)}\;\Delta^{2}_\psi \hat{H},
\end{equation}
where $\Delta^{2}_\psi\hat{H}:=\langle\psi|\hat{H}^{2}|\psi\rangle-\langle\psi|\hat{H}|\psi\rangle^{2}$ is now the variance of the Hamiltonian $\hat{H}$ evaluated for the state $|\psi\rangle$. Consequently, by just computing $g(0)$ for exemplary metrics discussed in \secref{sec:plethora}, we can directly write the scalar form for the QFI, WYD, Kubo-Mori and RLD metrics as:
\begin{align}
&\qquad \QFI\!\left(|\psi\rangle, \hat H\right)=\QFIg{\alpha=1/2}\!\left(|\psi\rangle, \hat H\right)
=4\Delta^{2}_\psi\hat{H} \nonumber\\
&\leq\quad
\QFIg{\alpha}\!\left(|\psi\rangle, \hat H\right)=\frac{1}{\alpha(1-\alpha)}\Delta^{2}_\psi\hat{H} \nonumber\\
\quad&\leq\quad
\KM\!\left(|\psi\rangle, \hat H\right)=\QFIR\!\left(|\psi\rangle, \hat H\right)=\infty,
\end{align}
which consistently respect the hierarchy \eref{eq:qmetric_hierachy} with the RLD and the Kubo-Mori informations (scalar metrics) being generally divergent on pure states.

On the other, \eqnref{eq:gmetric_unitary} takes a particular neat form in case of the \emph{WYD information} $\mathcal{I}_{\alpha}$---the scalar (single-parameter) version of the WYD divergence $\WYD_{\alpha}$ defined in \eqnref{eq:WYD_div}---which for $0\leq\alpha\leq1$ can be written as~\cite{Wigner1963,Lieb1973}:
\begin{equation}
\mathcal{I}_{\alpha}\left(\varrho, \hat H\right)=\frac{1}{2}\textrm{Tr}\left\{ \left[\varrho^{\alpha},H\right]\left[\varrho^{1-\alpha},H\right]\right\} ,
\end{equation}
and for the case of \emph{WY skew information} ($\alpha=1/2$) simplifies further to~\cite{Luo2004,Luo2004b,Luo2012}:
\begin{equation}
\mathcal{I}_{1/2}\!\left(\varrho, \hat H\right)=\frac{1}{2}\textrm{Tr}\left\{ \left[\sqrt{\varrho},H\right]^{2}\right\}.
\end{equation}

\begin{table*}[t]
\centering
\begin{tabular}
{ | M{3.75cm} | M{5cm}| M{1.5cm}| M{1cm}| M{3cm}|N }
\hline 
Quantum divergences & $D\!\left[\rho,\sigma\right]$ & $f(t)$ & $\ddot{f}(1)$ & Riemannian metric
&\\[10pt]
\hline 
\hline
(rescaled) Tsallis \eref{eq:Tsall_div}& $\frac{1}{\alpha}H_\alpha[\rho||\sigma]=\frac{1}{\alpha(1-\alpha)}\left(1-\xi_\alpha[\rho,\sigma]\right)$ & $\frac{1-t^{\alpha}}{\alpha(1-\alpha)}$ & $1$ & $\QFM{\alpha}$
&\\[8pt]
\hline 
R\'{e}nyi \eref{eq:qRenyi_div} & $D_{\alpha}[\rho||\sigma]=\frac{1}{\alpha-1}\ln\xi_\alpha[\rho,\sigma]$ & N/A & N/A & $\alpha\,\QFM{\alpha}$
&\\[8pt]
\hline  
Relative entropy \eref{eq:q_rel_entr}& $D[\rho||\sigma]=\textrm{Tr}\left[\rho\ln\rho-\rho\ln\sigma\right]$ & $-\ln t$ & $1$ & $\KM$
&\\[8pt]
\hline
Chernoff information \eref{eq:q_chernoff_info} & $C[\rho,\sigma]=-\ln(\min_{0\leq\alpha\leq1}\xi_\alpha[\rho,\sigma])$ & N/A & N/A & $\frac{1}{8}\,\QFM{\alpha=1/2}$
&\\[8pt]
\hline  
Bures distance & $B[\rho,\sigma]=2(1-F[\rho,\sigma])$ & N/A & N/A & $\frac{1}{2}\,\QFIM$
&\\[8pt]
\hline
Trace distance \eref{eq:trace_dist} & $T(\rho,\sigma)=\frac{1}{2}||\rho-\sigma||_{1}$ & $\frac{|1-t|}{2}$& N/A& N/A
&\\[8pt]
\hline
\end{tabular}
~\\~
\caption{Quantum divergences defined by means of quantum Chernoff coefficients $\xi_\alpha[\rho,\sigma]=\tr{\rho^\alpha \sigma^{1-\alpha}}$ and quantum fidelity $F[\rho,\sigma]=\left\Vert\sqrt{\rho}\sqrt{\sigma}\right\Vert_{1}\neq\xi_{1/2}[\rho,\sigma]$ interpreted as quantum $f$-divergences whenever possible, and their associated Riemannian metrics.}
\label{tab:div_quant}
\end{table*}

\subsection{Quantum metrics induced by divergences}
\label{subsec:Corresponding-divergence-induced}
In analogy with the classical information geometry, the quantum distance measures encountered in \secref{subsec:q_ht}, while considering the tasks of binary HT in the quantum setting, induce particular monotonic metrics onto the quantum statistical manifold $\cMq$. However, as in the previous section we have shown a plethora of such metrics to exist, it is important to identify by performing $\dparamV$-expansions as in \eqnref{eq:div_metric_q}, which of the $g$-metrics \eref{eq:g_metric} are induced by particular distinguishability measures.

Considering the sHT tasks of \secref{sec:sHT_quantum}, we expand first for infinitesimally close states $\rho_{\paramV}$ and $\rho_{\paramV+\dparamV}$ the quantum Chernoff coefficients \eref{eq:quant_Chernoff} and the fidelity \eref{eq:fidelity} to obtain:
\begin{align}
\xi_{\alpha}[\rho_{\paramV},\rho_{\paramV+\dparamV}]
&\approx
1-\frac{\alpha(1-\alpha)}{2}\dparamV^{T}\QFM{\alpha}(\rho_{\paramV})\dparamV,  \label{eq:xi_expansion}\\
F[\rho_{\paramV},\rho_{\paramV+\dparamV}]
&\approx 
1-\frac{1}{8}\dparamV^{T}\QFIM(\rho_{\paramV})\dparamV,
\label{eq:fidelity_expansion}
\end{align}
where the expansion \eref{eq:xi_expansion} follows from the fact that $\xi_\alpha=1-(1-\alpha)H_\alpha$ and the rescaled Tsallis divergence \eref{eq:Tsall_div}, $H_\alpha/\alpha$, we have shown in \secref{sub:WYD} to be the quantum $f$-divergence \eref{eq:q_f_div} inducing the WYD metric \eref{eq:WYD_metric}, $\QFM{\alpha}$. The expansion \eref{eq:fidelity_expansion} for fidelity can be obtained indirectly by just acknowledging that $F[\rho_,\sigma]=\cos(B_\trm{A}[\rho,\sigma])$ and recalling that the Bures angle $B_\trm{A}$ is the geodesic distance for the QFI metric \eref{eq:QFI_metric}~\cite{Sommers2003}. Furthermore, the expansion of the quantum Chernoff bound \eref{eq:quant_Chernoff} can just be deduced by minimising the expansion \eref{eq:xi_expansion} over $\alpha$, in order to obtain~\cite{Audenaert2007}:
\begin{equation}
\xi[\rho_{\paramV},\rho_{\paramV+\dparamV}]\approx1-\frac{1}{8}\dparamV^{T}\,\QFM{\alpha=1/2}\!\left[\rho_{\paramV}\right]\,\dparamV,
\label{eq:xi_q_infi}
\end{equation}
since the WYD metric \eref{eq:WYD_metric} is minimal for $\alpha=1/2$.

In case of aHT tasks of \secref{sec:aHT_quantum}, the quantum relative entropy \eref{eq:q_rel_entr} must expand to the Kubo-Mori metric \eref{eq:KM_metric} as follows
\begin{equation}
D[\rho_{\paramV}||\rho_{\paramV+\dparamV}]\approx\frac{1}{2}\dparamV^{T}\,\KM(\rho_{\paramV})\,\dparamV,
\label{eq:rel_entr_J}
\end{equation}
as $D[\rho||\sigma]$ constitutes a quantum $f$-divergence with $f(t)=-\ln t$ in \eqnref{eq:q_f_div}, so according to the relation \eref{eq:f_g} it must induce a $g$-metric with $g(t)=\frac{t-1}{\ln t}$ in \eqnref{eq:g_metric}, which is indeed the KM metric $\KM$. 

The expansion \eref{eq:rel_entr_J} of the quantum relative entropy is particularly useful when considering single-parameter perturbations of thermal states. A thermal state of a system, whose evolution is governed by a Hamiltonian $\hat{H}$, is given as $\omega=\ee^{-\hat{H}}/Z$, where $Z=\tr{\ee^{-\hat{H}}}$ can just be interpreted as a normalization constant. A small perturbation of the Hamiltonian $\hat{H}\to \hat{H}-\epsilon V$ with $\epsilon\ll1$ results in a disturbed thermal state $\omega_{\epsilon}=\ee^{-\hat{H}+\epsilon V}/Z\approx \ee^{-\hat{H}}/Z+\epsilon\,\delta\omega$, where $\delta\omega=\Omega_{\omega}\!\left[V\right]$ and the superoperator $\Omega_{\omega}$ is defined implicitly via the relation $\Omega_{\omega}^{-1}\left[A\right]=\left.\frac{\dd}{\dd t}\ln(\omega+tA)\right|_{t=0}$ holding for any Hermitian $A$~\cite{Beny2015}. The corresponding KM information (scalar KM metric \eref{eq:KM_metric}) for the $\epsilon$-perturbation can then be written as~\cite{Beny2015,DeBrota2016}:
\begin{equation}
\left.\KMI(\omega_\epsilon)\right|_{\epsilon=0}=\tr{\delta\omega V}=\tr{V\,\Omega_{\omega}\!\left[V\right]}.
\label{eq:KM_info}
\end{equation}
The above expression is especially appealing if $V=H$, i.e. the perturbation of the Hamiltonian occurs due to a change of temperature, in which case $\left.\KMI(\omega_\epsilon)\right|_{\epsilon=0}=\Delta^{2}_\omega\hat{H}$. In such a setting, the KM information equals just to the variance of the original Hamiltonian.

The above discussion is summarised in \tabref{tab:div_quant}, where we present all the quantum divergences and distances encountered in \secref{subsec:q_ht} together with the Riemannian metrics they induce. The crucial difference between the infinitesimal expansions of quantum divergences and their classical counterparts is that different distinguishability measures expand rather to different metrics in the quantum case. In particular, none of the above quantities except fidelity (or Bures distance) is directly related to the ordinary QFI metric \eref{eq:QFI_metric}, which can be interpreted as the most natural quantum generalisation of the classical Fisher metric \eref{eq:Fisher_metric} due to their common operational interpretation in the task of parameter estimation~\cite{BrandorffNielsen2000} (see also \secref{subsec:Quantum-metrology}). Strikingly, the quantum relative entropy that expands in \eqnref{eq:rel_entr_J} to the KM metric \eref{eq:KM_metric} is, unlike the QFI metric, even undefined on pure states. Note that, similarly to the classical TV distance \eref{eq:TV} presented in \tabref{tab:div_class}, the trace distance does not induce a Riemannian metric, as it corresponds to a quantum $f$-divergence with $f(t)=\frac{1}{2}|1-t|$ in \eqnref{eq:q_f_div} that is not differentiable at $t=1$.

\subsection{Relations between quantum divergences and metrics}

Quantum divergences, similarly to their classical counterparts, are related to each other trough various inequalities. Most of the relations mentioned in the classical setting in \secref{sec:class_relations} straightforwardly generalise to the quantum case. However, since the structure in the latter scenario is richer, some novel relation also arise~\cite{Audenaert2014}. As we did in the case of \secref{sec:class_relations}, in what follows we drop the quantum-state arguments of the quantum divergences considered whenever possible, in order to emphasise functional character of the relations.

\subsubsection{Relating quantum divergences}
Focussing on sHT tasks of \secref{sec:sHT_quantum}, the quantum Chernoff coefficients \eref{eq:quant_Chernoff}, $\xi_{\alpha}$, and the trace distance \eref{eq:trace_dist}, $T$, describing error in asymptotic and single-shot regimes may be related through the following "sandwich" inequality~\cite{Audenaert2008,Audenaert2014}:
\begin{equation}
\left.\begin{array}{c}
1-\xi_{\alpha}\\
1-F
\end{array}\right\} \leq T\leq\sqrt{1-F^{2}}\leq\sqrt{1-\xi_{1/2}^{2}},
\label{eq:TraceDist_Renyi}
\end{equation}
where $1-F\leq T\leq\sqrt{1-F^{2}}$ is often referred to as the  ``Fuchs-van der Graaf inequality''~\cite{Fuchs1999}, while the rightmost inequality is a simple consequence of the fact that $\xi_{1/2}\leq F$---affinity being always less than or equal to the fidelity \eref{eq:fidelity}. Note that although $1-\xi_{\alpha}$ and $1-F$ both serve as lower bounds on the trace distance, they are also interrelated via $F\leq\sqrt{\xi_{\alpha}}$~\cite{Audenaert2014}. This inequality can be proven by rewriting the fidelity \eref{eq:fidelity} as $F[\rho,\sigma]=||\rho^{1/2}\sigma^{1/2}||_1=||\rho^{(1-\alpha)/2}(\rho^{\alpha/2}\sigma^{(1-\alpha)/2})\sigma^{\alpha/2}||_1$ and applying the H{\"o}lder's inequality to the last expression.

Reversing the inequalities in \eqnref{eq:TraceDist_Renyi}, one obtains an equivalent inequality for the quantum Chernoff bound \eref{eq:quant_Chernoff}, $\xi$, as follows
\begin{equation}
\left.\begin{array}{c}
1-T\leq \xi_\alpha\\
F\leq \sqrt{1-T^2}
\end{array}\right\} \Longrightarrow 1-T\leq \xi \leq F\leq \sqrt{1-T^2},
\label{eq:ineq_qu}
\end{equation}
which should be interpreted as the quantum equivalent of the classical relation \eref{eq:class_chern_var} with an important novel middle inequality that arises due to affinity and fidelity constituting different distinguishability measures in the quantum setting, which generally obey $\xi_{1/2}\le F$.

In case of aHT, also the Pinsker inequality \eref{eq:clas_pinsker} has a direct quantum generalisation~\cite{Hiai1981}:
\begin{equation}
T[\rho,\sigma]\leq\sqrt{\frac{D\left[\rho||\sigma\right]}{2}},
\end{equation}
relating the trace distance \eref{eq:trace_dist} to the quantum relative entropy \eref{eq:q_rel_entr}. Moreover, one can derive an inequality completely analogous to \eqnref{eq:Pinsk_chern_clas} with the classical divergences replaced by their quantum counterparts. Similarly, by the same argument as in the classical case, inequality \eref{eq:relative_chernoff_class} also holds true in the quantum realm.

\subsubsection{Metric-induced bounds on the trace distance}

As mentioned already in \secref{subsec:Corresponding-divergence-induced} and \tabref{tab:div_quant}, the trace distance does induce a Riemannian metric on the quantum statistical manifold. Nonetheless, similarly to procedure for the TV distance discussed in \secref{subsec:Induced-relations-on}, we can perform the $\dparamV$-expansion of the trace distance $T[\rho_{\paramV},\rho_{\paramV+\delta\paramV}]$ in order to relate its dominating terms to the ones arising in expansions of other quantum divergences via inequalities \eref{eq:TraceDist_Renyi}. As in the classical case of \secref{subsec:Induced-relations-on}, we focus below only on the single-parameter case $\dparamV\equiv\delta\param$, leaving the generalization to higher dimensions to the reader.

Expanding in the inequalities \eref{eq:TraceDist_Renyi} to the lowest order in $\delta\param$ the quantum Chernoff coefficients $\xi_{\alpha}$ and the fidelity $F$ according to \eqnref{eq:xi_expansion}, one obtains
\begin{align}
\QFI(\rho_{\param})\frac{\delta\param^{2}}{8}
\leq&\;\mathcal{I}_{1/2}(\rho_{\param})\,\delta\param^{2} \\
&\leq\; 
T[\rho_{\param},\rho_{\param+\delta\param}] \nonumber\\
&\quad\leq\;
\sqrt{\QFI(\rho_{\param})}\,\frac{\delta\param}{2}\leq\sqrt{2\mathcal{I}_{1/2}(\rho_{\param})}\,\delta\param, \nonumber
\end{align}
where the inequalities containing WYD information apply above \emph{only if} the eigenvalues of the state $\rho_\param$ do not change with $\param$. Upon taking the tightest bounds from both sides, one arrives at
\begin{equation}
\mathcal{I}_{1/2}\left[\rho_{\param}\right]\delta\param^{2}\leq T[\rho_{\param},\rho_{\param+\delta\param}]\leq\sqrt{\mathcal{\FI_{\textrm{Q}}}\,\left[\rho_{\param}\right]}\frac{\delta\param}{2},
\label{eq:last_ineq}
\end{equation}
where the trace distance can be further expanded by considering two infinitesimally close states $\rho_\param$ and $\rho_{\param+\delta\param}=\rho_\param+\left.\frac{\partial \rho}{\partial \param}\right|_\param \delta\param+O(\delta\param^2)$, as follows
\begin{equation}
T[\rho_{\param},\rho_{\param+\delta\param}]=\frac{1}{2}||\rho_{\param}-\rho_{\param+\delta\param}||_{1}=\frac{1}{2}\left\Vert \frac{\partial\rho_{\param}}{\partial\param}\right\Vert _{1}\delta\param +O(\delta \param^2).
\end{equation}
Hence, it is the rightmost inequality in \eqnref{eq:last_ineq} that correctly recovers the dominating behaviour of the trace distance as $\delta\param\to0$, yielding an inequality
\begin{equation}
\left\Vert \frac{\partial\rho_{\param}}{\partial\param}\right\Vert _{1}^{2}\le\FI_{\textrm{Q}}\!\left[\rho_{\param}\right]
\label{eq:derivative_QFI}
\end{equation}
that holds in general for any $\rho_\param$---previously reported for the case of unitary encodings~\cite{Oszmaniec2016}. Moreover, in the context of coherence theory, the trace norm of state derivative is referred as the \emph{asymmetry} and used to quantify coherence of a given state~\cite{Marvian2014}.

\section{Applications}
\label{sec:Applications}

\subsection{Quantum parameter estimation\label{subsec:Quantum-metrology}}
A complementary statistical inference task to HT discussed in \secref{subsec:Hypothesis-testing} is the problem of \emph{parameter estimation}. Let us consider a family of PDs $p_\param(x)$ that is parametrised by a single parameter $\param$ and describes results of an experiment depending on an unknown value of $\param$. The main goal is to find the true value of $\param$, which can be also seen as identification of a correct probability distribution describing the system. However, in contrast to HT, one deals here with a continuous family of PDs. The value of $\param$ is estimated using an estimator function $\tilde{\param}(x)$ which provides a guess about the true value of the parameter. The precision of the estimation procedure is quantified by means of the a \emph{mean squared error} (MSE), defined as $\Delta^{2}\param:=\sum_{x}p_\param(x)\left(\tilde{\param}(x)-\param\right)^{2}$ which must be minimised for the best performance. For unbiased estimators, that is for estimators that on average return the true value of the parameter, i.e.~$\sum_{x}p_\param(x)\tilde{\param}(x)=\param$, the \emph{Cram\'{e}r-Rao inequality}~\cite{Cover1991}:
\begin{equation}
\label{eq:CRB}
\Delta^{2}\param\ge\frac{1}{\nu\,\FI\!(p_\param)},
\end{equation}
provides a general lower bound on the MSE. Here, $\nu$ is the number of independent repetitions of the experiment and $\FI(p_\param)$ is nothing but the Fisher information (FI) defined in \eqnref{eq:FI}. Importantly, the Cram\'{e}r-Rao bound \eref{eq:CRB} is guaranteed to be tight in the $\nu\to\infty$ limit of large number of repetitions, e.g.~by the maximum likelihood estimator~\cite{Lehmann1998}, what, on the other hand, gives a clear operational meaning to the FI.

In the quantum version of the parameter estimation task one deals with a parameter-dependent quantum state $\rho_\param$ to be measured by some POVM $\{\Pi_{x}\}_x$, whose outcome
$x$ occurs then with probability $p_\param(x)=\textrm{Tr}\left\{ \rho_\param\Pi_{x}\right\}$ and is used to estimate the true value of $\param$. Since idealistically the experimentalist has control over the measurement apparatus, she can choose the optimal detection scheme giving the best estimation precision for a given state $\rho_\param$. Hence, in order to obtain the corresponding tightest quantum version of the bound \eref{eq:CRB} on precision. one must maximise the FI appearing in \eqnref{eq:CRB} over all the possible measurements. Although such procedure may seem to be difficult, it can be shown that it results in the quantum Fisher Information (QFI) $\QFI(\rho_\param)$~\cite{Braunstein1994,BrandorffNielsen2000}, i.e.~the scalar version of the QFI metric \eref{eq:QFI_metric}, that bounds then the precision through the \emph{quantum Cramer-Rao inequality}~\cite{Helstrom1976}:
\begin{equation}
\Delta^{2}\param\geq\frac{1}{\nu\,\QFI(\rho_\param)},
\label{eq:QCRB}
\end{equation}
which is guaranteed to coincide with the bound \eref{eq:CRB} in the asymptotic $\nu$ limit, when one performs a identical projective measurements on each copy of the state in the eigenbasis of the SLD operator (see \eqnref{eq:QFI_metric}) for the $\param$-parameter~\cite{Braunstein1994} and employs the maximum likelihood estimator.

\subsection{Quantum speed limits}

An interesting problem in which quantum metrics play a crucial role is the determination of the so-called \emph{quantum speed limits}. The task is to derive a lower bound on the minimal time $\tau$ required for a given evolution specified by a quantum map $\Lambda_{t}$ to drive the system from an initial state $\rho_{t=0}=\rho$ to a final state $\rho_{t=\tau}=\sigma$~\cite{Mandelstam1945,Margolus1998,Campo2013,Taddei2013}. Intuitively, states that are poorly distinguishable lie close to each other and therefore $\tau$ should be small, while easily distinguishable should require a larger $\tau$ to be evolved onto one another. In order to formalise this notion, let us choose a particular metric $\g$ on the quantum statistical manifold $\cMq$ and consider a curve $\rho_{t}=\Lambda_{t}[\rho]$ parametrised by $t\in[0,\tau]$. The length of such a curve can be evaluated as $S(\tau)=\int_{0}^{\tau}\sqrt{\sum_{ij}\g(\rho_t)_{ij}\frac{\dd e_{i}}{\dd t}\frac{\dd e_{j}}{ \dd t}} \dd t=\int_{0}^{\tau}\sqrt{\g(\rho_t)_{tt}} \dd t$, where $\{e_{i}\}_i$ are the basis vectors of $\tT_{\rho_t}\cMq$ at a given $\rho_t\in\cMq$ that can always, in principle, be chosen so that only one of them points along the $t$-curve. One can, however, take a different path on $\cMq$, in particular, choose the shortest one as defined by the geodesic length between the states $\rho$ and $\sigma$, denoted as $\mathcal{L}_\g[\rho,\sigma]$. Hence, one can in general write
\begin{equation}
\frac{1}{\tau}\mathcal{L}_\g[\rho,\sigma]\leq\frac{1}{\tau}\int_{0}^{\tau}\sqrt{\g_{tt}}\dd t
\quad\implies\quad
\tau\geq\frac{\mathcal{L}_\g[\rho,\sigma]}{\overline{\sqrt{\g}}},\label{eq:QSL_general}
\end{equation}
where $\overline{\sqrt{\g}}:=\frac{1}{\tau}\int_{0}^{\tau}\sqrt{\g_{tt}}\dd t$ can then be interpreted as the square root of the metric averaged over time. Interestingly, for various choices of the metric $\g$ above one gets different quantum speed limits. In particular, it was shown that for some quantum dynamics between states the quantum speed limit \eref{eq:QSL_general} obtained with help of the WYD metric \eref{eq:WYD_metric} may turn out to be tighter then the one originating from the QFI metric \eref{eq:QFI_metric}~\cite{Pires2016}, despite the latter metric constituting in fact the minimal one in the hierarchy of monotonic metrics \eref{eq:qmetric_hierachy}.

\subsection{Quantum thermodynamics\label{subsec:Quantum-thermodynamics}}

Classical thermodynamics is usually concerned with large systems, whose macroscopic properties can be inferred in the thermodynamic limit without resorting to the detailed microscopic description of the system. Obviously this assumption is not satisfied if one wants to delve into the micro-world governed by the laws of quantum mechanics. It is therefore interesting how the classical laws of thermodynamics must be generalised or modified for them to also apply when dealing with quantum systems---a question that has recently motivated a rapid development in the field of \emph{quantum thermodynamics}~\cite{Deffner2019}.

Let us consider a process in which a system is governed by a Hamiltonian $\hat{H}$, while initially being prepared in a thermal state $\omega_{0}=e^{-\beta H}/Z$ at a temperature $\beta$. Given that it evolves into a state $\omega_{1}$, the quantum relative entropy \eref{eq:q_rel_entr} between the final and initial states reads
\begin{equation}
\label{eq:rel_entr_thermo}
D\left[\omega_{1}||\omega_{0}\right]=\beta\left(F(\omega_{1})-F(\omega_{0})\right)=\beta\,\Delta\langle \hat H\rangle-\Delta S,
\end{equation}
where $F(\omega):=\tr{\omega\hat{H}}-S(\omega)/\beta$ is the free energy of state $\omega$, whereas $\Delta S:=S(\omega_{0})-S(\omega_{1})$ and $\Delta\langle \hat{H}\rangle:=\tr{\omega_{1} \hat H}-\tr{\omega_{0}\hat H}$. In particular, the quantum relative entropy is effectively equal to the difference between the free energies of the final and initial states. Now, since the relative entropy constitutes a non-negative quantum $f$-divergence \eref{eq:q_f_div}, it follows that $\Delta S\leq\beta\,\Delta\langle \hat H\rangle$, which is just the familiar \emph{Clausius inequality}. 

Let us assume that the process $\omega_{0}\to\omega_{1}$ is parametrised by a parameter $\lambda\in\left[0,\delta\lambda\right]$ and the final and initial states are close to each other, so $\delta\lambda\ll1$. Then, one may rewrite \eqnref{eq:rel_entr_thermo} by Taylor-expanding the relative entropy in $\delta\lambda$, as in \eqnref{eq:rel_entr_J}. Although to first-order $O(\delta\lambda)$ one obtains then $\Delta S=\beta\,\Delta\langle \hat{H}\rangle$, as the expansion \eref{eq:rel_entr_J} of the quantum relative entropy must always start at $\Theta(\delta\lambda^2)$, the leading term by which $\beta\,\Delta\langle \hat H\rangle$ is greater than $\Delta S$ is then given by $\delta\lambda^{2}\,\KMI(\omega_0)/2$~\cite{Deffner2010}. In other words, one may say that the KM information \eref{eq:KM_info} dictates the amount by which the equality in $\Delta S\leq\beta\Delta\langle H\rangle$ is violated.

\section{Conclusions}

We have reviewed relations between various measures of statistical distances and divergences from the perspective of both the classical and quantum statistical inference tasks. In particular, we have focussed on interrelating them, while studying what geometric structures they impose on the classical and quantum statistical manifolds in the language of information geometry. The quantum setting is much richer, especially because different quantum divergences induce distinct quantum metrics on the quantum statistical manifold in stark contrast to the classical case---even when restricting only to the operationally motivated ones that are monotonic under the action of quantum maps. This, however, creates an interesting playground of quantum monotonic metrics that then naturally arise in different forms in problems of quantum information theory, such as:~quantum parameter estimation, quantum speed limits or quantum thermodynamics.

\begin{acknowledgments}
This work was supported by the Foundation for Polish Science within the ``Quantum Optical Technologies'' project carried out within the International Research Agendas programme cofinanced by the European Union under the European Regional Development Fund.
\end{acknowledgments}

\bibliography{geom_infer}

\end{document}